\documentclass[journal]{IEEEtranTIE}
\usepackage{amsmath,amsfonts}
\usepackage{algorithmic}
\usepackage{algorithm}
\usepackage{array}
\usepackage[caption=false,font=normalsize,labelfont=sf,textfont=sf]{subfig}
\usepackage{textcomp}
\usepackage{stfloats}
\usepackage{url}
\usepackage{verbatim}
\usepackage{cite}
\usepackage{multicol}
\usepackage{graphics} 
\usepackage{epsfig} 
\usepackage{mathptmx} 
\usepackage{times} 
\usepackage{amsmath} 
\usepackage{amssymb}  
\usepackage{graphicx}
\usepackage{relsize}
\DeclareMathAlphabet{\mathcal}{OMS}{cmsy}{m}{n}
\SetMathAlphabet{\mathcal}{bold}{OMS}{cmsy}{b}{n}
\graphicspath{{./fig/}}
\usepackage[usenames,dvipsnames]{color}
\usepackage{epstopdf}
\usepackage{caption}

\usepackage{amsmath,amsthm,amssymb,mathrsfs,dsfont}
\usepackage{amsfonts}
\usepackage{siunitx}
\usepackage[colorinlistoftodos,prependcaption,textsize=tiny]{todonotes}

\usepackage{tikz}
\usetikzlibrary{shapes.geometric, arrows.meta, positioning}
\usepackage{graphicx}
\usepackage{cite}
\usepackage{picinpar}
\usepackage{amsmath}
\usepackage{url}
\usepackage{flushend}
\usepackage[latin1]{inputenc}
\usepackage{colortbl}
\usepackage{soul}
\usepackage{multirow}
\usepackage{pifont}
\usepackage{color}
\usepackage{alltt}
\usepackage[hidelinks]{hyperref}
\usepackage{enumerate}
\usepackage{siunitx}
\usepackage{breakurl}
\usepackage{epstopdf}
\usepackage{pbox}

\usepackage{verbatim} 

\newtheorem{assumption}[]{Assumption}

\usepackage{tikz}
\usetikzlibrary{shapes.geometric, arrows.meta, positioning}

\usepackage[cuteinductors]{circuitikz}
\tikzset{%
	>={Latex[width=2mm,length=2mm]},
	base/.style = {rectangle, rounded corners, draw=black,
		minimum width=1.5cm, minimum height=0.5cm
	},
	box/.style = {base, minimum width=2cm,text centered, minimum height= 1.1cm},
	textbox/.style = {base, minimum width=1cm,align= center }
}

\definecolor{forestgreen}{rgb}{0.13, 0.55, 0.13}
\definecolor{darkblue}{rgb}{0,0,0.8}
\definecolor{maigreen}{rgb}{0.549, 0.714, 0.235}

\definecolor{rev1}{rgb}{0,0,0}
\definecolor{rev2}{rgb}{0,0,0}

\usepackage{cite}

\renewcommand{\baselinestretch}{0.957}

\newcommand\oprocendsymbol{\hbox{$\blacksquare$}}

\newcommand{\dd}[0]{\mathrm d}
\newcommand\oprocend{\relax\ifmmode\else\unskip\hfill\fi\oprocendsymbol}

\newcommand{\real}[0]{\mathbb R}

\providecommand{\norm}[1]{\lVert#1\rVert}

\newcommand{\com}[1]{}

\usepackage[normalem]{ulem}
\newcommand\rout{\bgroup\markoverwith{\textcolor{red}{/}}\ULon} 

\usepackage[prependcaption,colorinlistoftodos]{todonotes}

\expandafter\def\expandafter\normalsize\expandafter{%
	\normalsize%
	\setlength\abovedisplayskip{1pt}%
	\setlength\belowdisplayskip{2pt}%
	\setlength\abovedisplayshortskip{-8pt}%
	\setlength\belowdisplayshortskip{2pt}%
}

\begin{document}
\title{Hardware test and validation of the angular droop control: Analysis and experiments$^*$}

\author{
	\vskip 1em
	
\emph{Taouba Jouini$^{1,2}$},
	\emph{Jan Wachter$^{2}$}, \emph{Sophie An$^{2}$ and Veit Hagenmeyer$^{2}$ }
	\thanks{
	$^*$ The first two authors contributed equally.
	
	$^{1}\,$Institute of Automatic Control, Leibniz University of Hannover
	
	$^{2}\,$Institute for Automation and Applied Informatics at Karlsruhe Institute of Technology, Germany. 
	Emails: \{\tt jouini@irt.uni-hannover.de, jan.wachter@kit.edu, sophie.an@alumni.kit.edu, veit.hagenmeyer@kit.edu.\}
}
	}

\maketitle
\begin{abstract}
We present a hardware-based validation of angular droop control for grid-forming DC/AC converters, a control strategy that establishes active power-to-angle droop.
Angular droop control enables exact frequency regulation at steady state, thereby combining primary and secondary control into a single layer.
We provide traceable analysis and suggest solutions to the main implementation challenges with angular droop control, specifically addressing the challenges concerning discretization and clock drift in hardware experiments.
This is illustrated in two different scenarios.
Experimental results from the single converter to load scenario demonstrate black start capability and power-to-angle droop behavior for two different implementation schemes.
A multi-converter setup validates frequency synchronization and power-sharing properties, proving the ancillary services that angular droop control provides in the real-world experimental setup.
\end{abstract}

\begin{IEEEkeywords}
Hardware experiments, DC/AC power converters, Grid-forming inverters, Power systems, Control.
\end{IEEEkeywords}

\markboth{}%
{}

\definecolor{limegreen}{rgb}{0.2, 0.8, 0.2}
\definecolor{forestgreen}{rgb}{0.13, 0.55, 0.13}
\definecolor{greenhtml}{rgb}{0.0, 0.5, 0.0}

\section{Introduction}

\IEEEPARstart{P}{ower} grids are facing a rapid transition from fossil fuel towards an increasing share of renewable energy resources.
This profound change is characterized by the integration of converter-based generation~\cite{sajadi2023}.
In particular, the high penetration of power electronics alters the power system dynamics, governed thus far by rotating synchronous machines underpinning the legacy grid~\cite{wachter2024}. Therefore, the control of DC/AC converters lies at the forefront of this transition to ensure power system stability~\cite{dorfler2023a}.

Typical control approaches for grid-forming DC/AC converters are inspired by the dynamics governing synchronous machines and their analogy to coupled oscillator dynamics~\cite{SIMPSONPORCO20132603}.
For example, {\em frequency droop} control is based on the active power to frequency droop, which is inherent to synchronous machines and enables the synchronization between different generators.
This grid-forming method is extensively studied both in theory and practice.
An experimental validation of the frequency droop control strategy for different applications can be found in~\cite{chakraborty2024}.
The dynamics of synchronous machines remain a source of inspiration for a multitude of converter control strategies that emulate their behavior.
One particular controller that relies on exact model matching of high-order dynamics of synchronous machines is the {\em matching control} introduced in~\cite{JOUINI2016192,ARGHIR2018273}.
This controller relies on easily measured DC-side voltage to play the role of an indicator of power imbalance in the grid.
Oscillator-based control schemes such as the {\em Virtual Oscillator Control}~\cite{6692879} rely on emulating the dynamics of weakly coupled nonlinear oscillators.
Compared to frequency droop control, which is only well-defined in the vicinity of a sinusoidal steady state, virtual oscillator control enables interconnected converters to stabilize synchronous sinusoidal waveforms starting from arbitrary initial condition.
The virtual oscillator control is validated experimentally in~\cite{johnson2015synthesizing} within a laboratory hardware prototype to demonstrate the validity of the design approach.
Even though the virtual oscillator control has provable active power to frequency droop properties~\cite{7171084}, the control tuning remains a difficult task due to a lack of intuition on the physical interpretation of the gains.
It was also not possible to track active and reactive power setpoints in its original formulation in~\cite{6692879}.
These limitations have motivated a variant of Virtual Oscillator Control suggested in~\cite{8638531} that allows active and reactive power to be dispatched, hence the name {\em dispatchable Virtual Oscillator Control} whose experimental validation is conducted in~\cite{seo2019dispatchable}.
\textcolor{rev1}{The authors in~\cite{Strunk2024} design an angle based modification for a frequency droop controller to dampen power oscillations in a multiple converter setup.}

All the control approaches discussed above have in common that they are based on the principle of active power to frequency droop.
In contrast, the \textit{angular droop} control studied in this paper establishes a linear relationship between active power and angle deviation, rather than frequency.
This achieves exact frequency regulation at steady state.
As a consequence, it merges primary with secondary frequency control~\cite{5275987} and no additional control layer, i.e., secondary control is necessary.
During transients, compared to frequency droop control, the angular droop acts on the rate of change of frequency to counterbalance the rate of change of power which anticipates a change in the power balance itself.
Therefore angular droop control reacts faster to load disturbances.
Additionally, the angular droop control is shown to be inverse optimal stabilizing for the angle dynamics~\cite{jouini2022inverse,tj_phd}.
The optimality of the angular droop control brings about inherent desirable gain margins analogous to linear quadratic regulators and showcases the utility of inverse optimal control theory
in networked settings~\cite{9429728}.
\textcolor{rev1}{Angular droop control has been tested in simulations on numerous power system benchmarks. In \cite{5275987} the operation of a simulated microgrid consisting of angular droop controlled DC/AC converters is presented and the zero frequency deviation property at steady state is shown in comparison to frequency droop control. Kolluri et. al~\cite{kolluri_2017} examine the power sharing properties of angular droop control and validate their results using a simulation study. Xu et. al~\cite{Xu2021} deploy a simulation study of an islanded microgrid to validate an angular droop based method for cost minimization and consensus active power sharing between distributed generation units. In \cite{jouini2022inverse} it is proven that angular droop is an inverse optimal locally stabilizing control law for a multi-converter system, which is supported by numerical simulations. Those results are extended~\cite{jouini2023a} to account for input and output constraints posed by real-world applications and the tuning of discrete-time implementations of angular droop control is discussed in~\cite{jouini2023b}. Both studies support their results using numerical simulations.}

The sole deployment of numerical simulations for validation is unsatisfactory due to the discrepancy between real-world setups and simplified settings adopted in numerical case studies such as unmodeled dynamics and erroneous or unknown model parameters affecting the system.

This work demonstrates the grid-forming properties of angular droop control in a controlled experimental setting~\cite{wiegel2022smart}, with particular emphasis on the hardware-based validation. The main contributions are as follows:
\begin{itemize}
    \item This paper presents the first hardware-based implementation of angular droop control. The challenges of opting for angle- instead of frequency-based control are addressed in a real hardware experimental setup.
    \item Grid-forming properties such as black start, robustness to load changes, and zero steady state frequency deviation property of angular droop control are validated using the single converter-to-load scenario. The control law is reformulated to suit hardware realization, and two implementation schemes are compared to show compatibility with different inner control architectures.
    \item Frequency synchronization and power-sharing capabilities are proven in the two-converter scenario, forming the basis for generalization to $n$-converter setups. Clock drift issues are analyzed and mitigated via master clock distribution. Further, practical guidance on control tuning for angular droop is provided.
\end{itemize}
This experimental validation represents a meaningful contribution toward the practical deployment of angular droop control, helping to bridge the gap between theoretical development and real-world implementation.

The paper is structured as follows. Section~\ref{sec:Understand-adroop-ctrl} first briefly introduces angular droop control and discusses the difference to frequency droop control, then details the experimental environment.
Thereafter, in Section~\ref{sec:scenario-I} the results of the single converter to load scenario are discussed along with the challenge of discretizing the angle dynamics.
Section~\ref{sec:scenario-II} is concerned with the two-converter to load scenario and the challenges connected to synchronization and power-sharing.
Lastly, Section~\ref{sec:concl} summarizes the findings and concludes the paper.

{\em Notation:} 
Define $\mathbf{I}=\left[\begin{smallmatrix}
	1 &0 \\ 0 & 1 
\end{smallmatrix}\right]$ and $\mathbf{J}=\left[\begin{smallmatrix}
	0 &-1 \\ 1 & 0 
\end{smallmatrix}\right].$
Let $x$ denote an AC quantity in $abc-$frame and $x_{dq}:=\mathcal{P}(\theta_{dq})\,  x$ denote its transformation in $dq-$frame~\cite{kundur1994power} following a Park transformation $\mathcal{P}(\theta_{dq})$ with angle $\theta_{dq}(t):=\theta_k(t)$, where $\theta_k$ is the angle of converter $k$. \textcolor{rev2}{Consider a network described by a connected graph~$\mathcal{G}=(\mathcal{V}, \mathcal{E},  \Xi)$, consisting of $\vert\mathcal{V}\vert=n$ nodes representing  DC/AC converter buses and $\vert\mathcal{E}\vert=m$ edges modeling purely inductive transmission lines with susceptance $b_{kj}>0,\; (k,j)\in\mathcal{E}$ collected in the diagonal matrix~$\Xi=\mathrm{diag}(b_{kj}),\; (k,j)\in\mathcal{E}$. The topology of the graph~$\mathcal{G}$ is described by the incidence matrix $\mathcal{B}\in\real^{n\times m}$.}
\section{The angular droop control}
\label{sec:Understand-adroop-ctrl}
\subsection{Control scheme}
\label{subsec: ctrl-tuning}
Consider a network of DC/AC converters, each represented by a voltage phasor. Hereby all the phasors are modeled with a constant magnitude (e.g. one per unit), and the converter's angle dynamics are assumed to be controllable. Overall the network dynamics can be represented by,
\begin{align}
	\label{eq: conv-dyn}
	\dot\theta= \hat u\,(\theta)+\omega^*\mathds{1}_n,\quad \theta(0)=\theta_0, 
\end{align}
where $\hat u(\theta)=[\hat u_1(\theta),\dots,\hat u_n(\theta)]^\top\in\mathbb{R}^n$ is the main control input, $\theta=[\theta_1,\dots,\theta_n]^\top\in\mathbb{R}^n$ is the vector of phase angles of the DC/AC converters, $\theta_0=[\theta_{0,1},\dots,\theta_{0,n}]^\top\in\mathbb{R}^n$ is the initial angle vector and $\omega^*$ is the nominal angular frequency. The angular droop control is given by~\cite{jouini2022inverse,tj_phd},
\begin{align}
	\label{eq: angular droop}
	\hat u(\theta) =  -\frac{1}{2}R^{-1}\, (\Gamma\, (\theta- \theta^*)+ P(\theta)-P^*),
\end{align}
with $R=\text{diag}(\alpha_1, \dots, \alpha_n)>0, \, \Gamma=\text{diag}(\gamma_1, \dots, \gamma_n)>0$. Further, $P(\theta) = [P_1 (\theta),\dots,P_n(\theta)]^\top\in\mathbb{R}^n$ denotes the active power vector, and $\theta^*=[\theta^*_1,\dots,\theta^*_n]^\top\in\mathbb{R}^n$ and $P^* =[P^*_1 (\theta),\dots,P^*_n(\theta)]^\top\in\mathbb{R}^n$ the respective reference values.
Note that for the angular droop control design \eqref{eq: angular droop}, we assume that synchrophasor measurements of the angle, with respect to a global frame of reference, are available to each converter. This is a reasonable scenario for a future power grid, as phasor measurement unit (PMU) installation is becoming increasingly widespread\textcolor{rev2}{\cite{Golshani2021,wang2020iet,Fank2024}}.
In summary, the closed-loop angle dynamics are given by,
\begin{align}
	\label{eq: cl-dyn}
	\dot\theta = -\frac{1}{2\alpha} \big(\gamma\, (\theta- \theta^*)+ P(\theta)-P^*)+\omega^*\mathds{1}_n,
\end{align} 
where the gain matrices $R = \alpha\,  \mathbf{I}_n$ and $\Gamma= \gamma\, \mathbf{I}_n$ with $\alpha, \gamma>0$, i.e., the control gains are uniform across all the converters. 
Observe that:
\begin{itemize}
	\item A decrease in the gain $\alpha>0$ improves the angle transients, i.e., it results in faster convergence of the angles towards the induced steady state angle. 
	\item The gain $\gamma>0$ defines the power-to-angle droop behavior between the power and angle deviation at steady state characterized by
	\begin{align}
		\label{eq: droop-bhv}
		\Gamma\, (\theta^s -\theta^*)  =  P^*-P^s(\theta^s),
	\end{align} 
	where $\theta^s\in\real^n$  and $P^s(\theta^s)$ are the vectors of induced steady state angles and powers, respectively.
	\item Both the gains $\gamma$ and $\alpha$ affect the rate of change of frequency or RoCoF given by $\dot\omega$. This can be seen by taking the time derivative of~\eqref{eq: cl-dyn} as follows
	\begin{align}
		\ddot\theta = -\frac{1}{2\alpha} \big(\gamma (\dot\theta-\dot\theta^*)+ \dot P(\theta)\big),
	\end{align}
	where $\dot P(\theta)=\frac{\dd P(\theta)}{\dd t}$. By letting $\omega := \dot\theta$ and $\omega^* := \dot\theta^*$, we have
	\begin{align}
		\label{eq: rocof}
		\dot\omega = -\frac{1}{2\alpha} \big(\gamma\,  (\omega-\omega^*)+ \dot P(\theta)\big).
	\end{align}
	Therefore, a sudden change in active power corresponds to a sudden RoCoF $\dot{\omega}$ that occurs during the transients, while the frequency error remains zero at steady state. Observe that the RoCoF depends on both gains $\alpha$ and $\gamma$.
\end{itemize}
\color{rev2}
\subsubsection*{Stability and optimality \cite{jouini2022inverse}}
\label{subsec: stab-opt}
For clarity of exposition, we introduce the error coordinates $\tilde \theta(t)=\theta(t)-\theta^*(t)$, the angle vector at induced steady state $\theta^s := \lim_{t\to\infty} \theta(t)$ and consider the following optimal control problem,
\begin{align}
	\label{eq: main-prob}
	\min_{u\in\real^n}& \int_0^\infty \sum_{k=1}^n \bigg(\alpha_k  u_k^2(\tilde\theta)+ \frac{1}{4\alpha_k}\Big(\gamma_{k}\tilde\theta_k+ P_{k}(\tilde \theta)- P^*_{k}\Big)^2\bigg)\, \mathrm{d}t, \nonumber \\
	&\text{s.t. } \dot{\tilde\theta} = \hat u(\tilde\theta),\quad \tilde\theta(0)=\tilde\theta_0.
\end{align}

\begin{assumption}\cite{jouini2022inverse}
	\label{ass: bounded-sol}
	The induced steady state angle vector $\tilde\theta^s=\{\tilde\theta^s_k\}_{k=1}^n$ satisfies, 
	$\mathcal{B}^\top \tilde\theta^{ s}\in\left(-\frac{\pi}{2}, \frac{\pi}{2}\right)^m$, where $\mathcal{B}\in\real^{n\times m}$ is the incidence matrix of the underlying graph $\mathcal{G}$.
\end{assumption}

Under Assumption~\ref{ass: bounded-sol}, the angular droop control is the optimal stabilizing solution of~\eqref{eq: main-prob} in a neighborhood of the induced steady state angle $\tilde \theta^s$ that satisfies
\begin{align}
	\Gamma\, \tilde \theta^s  =  P^*-P^s(\tilde \theta^s),
\end{align}
where $P^s(\tilde \theta^s)=[P_{1}^s(\tilde \theta^s),\dots, P_{n}^s(\tilde \theta^s)]^\top$ is vector of the induced steady state active power~\cite{jouini2022inverse}  . 
Note that~\eqref{eq: droop-bhv} describes the steady state as a power balance between the active power and angle deviation from the nominal value. It coincides with the steady state resulting from letting the running cost in~\eqref{eq: main-prob} go asymptotically to zero, i.e., 
\begin{align*}
    \lim\limits_{t\to \infty}\left(\gamma_{k}\tilde\theta_k(t)+  P_{k}(\tilde \theta(t))-P^*_{k}\right)=0 ,\quad \forall\, k = 1,\dots, n.
\end{align*}
\color{black}
\subsection{Angular vs. frequency droop -- Distinction of properties}
\label{sec:angle_frequency_droop_comp}
In this section, we highlight the merits of the angular droop control compared to frequency droop control in terms of frequency support and steady state behavior.
In particular, we provide a comprehensive analysis that demonstrates why angular droop advances the state-of-the art. 

Inspired by the dynamics governing synchronous machines and their analogy to coupled oscillator dynamics~\cite{SIMPSONPORCO20132603}, frequency droop control is a grid-forming method that obeys the following closed-loop dynamics:
	\begin{align}
		\label{eq: freq-droop-dyn}
		\dot\omega &= -\frac{1}{2 \alpha} \big(\gamma\, (\omega- \omega^*)+ P(\theta)-P^*)\,. 
	\end{align}
For easiness of comparison, we hereby denote the virtual inertia  $M = 2\, \alpha\, \mathbf{I}_n$ and the damping coefficient $D = \gamma\, \mathbf{I}_n$.

Linking active power to angle instead of frequency yields altogether different behavior during both transient and steady state, by comparing the frequency droop dynamics~\eqref{eq: freq-droop-dyn}  to angular droop dynamics~\eqref{eq: cl-dyn}, we observe that:
\paragraph{Frequency support}
For the angular droop control, the rate of change of frequency or RoCoF is induced by the rate of change of power in~\eqref{eq: rocof} and not the power imbalance itself as in~\eqref{eq: freq-droop-dyn}. This means that the angular droop acts on the rate of change of frequency to counterbalance the rate of change of power which anticipates a change in the power balance itself.
This is analogue to a derivative control that extrapolates the current slope of the error and generates one large corrective effort immediately after a load change in order to begin eliminating the error as quickly as possible.
A loop with derivative control recovers quicker from a disturbance with less deviation than a loop with only a proportional term as in frequency droop control.
\paragraph{Steady state behavior}
At steady state, the frequency droop behavior established by~\eqref{eq: freq-droop-dyn} is given by 
    \begin{align}
		\gamma \, (\omega^s- \omega^*) = P^*-P(\theta^s).
    \end{align}
Here, the induced steady state frequency $\omega^s$ deviates from the nominal frequency $\omega^*$ and depends directly on the magnitude of the disturbance affecting the power network.
Therefore a secondary control is necessary to bring the steady state frequency back to nominal.
On the other hand, the steady state frequency induced by angular droop control, see \eqref{eq: droop-bhv}, settles to its nominal value at steady state in disregard of the disturbance amplitude implying {\em zero} frequency error and thus merging primary and secondary control~\cite{5275987}. \textcolor{rev1}{Please refer to Fig.~\ref{fig:angular_vs_frequency_droop_loadstep} in Sec.~\ref{sec:experimentalresults} for the experimental results on this comparison}. This saves additional control effort while achieving the same result.

\subsection{Experimental setup}
The hardware testbed consists of programmable DC/AC converter systems, a resistive load and transmission line replicas. Relevant parameters of the setup are summarized in Table~\ref{tab: dcac_param}. Details about the model under consideration are presented in Appendix~\ref{ap:appendixA}. \textcolor{rev1}{For further information on the microgrid laboratory refer to \cite{wiegel2022smart}.}

\begin{figure}[ht]
	\centering
	\begin{tikzpicture}[align=center]
		\node[inner sep=0pt] (imperix) at (0,0) {\includegraphics[width=0.3\textwidth]{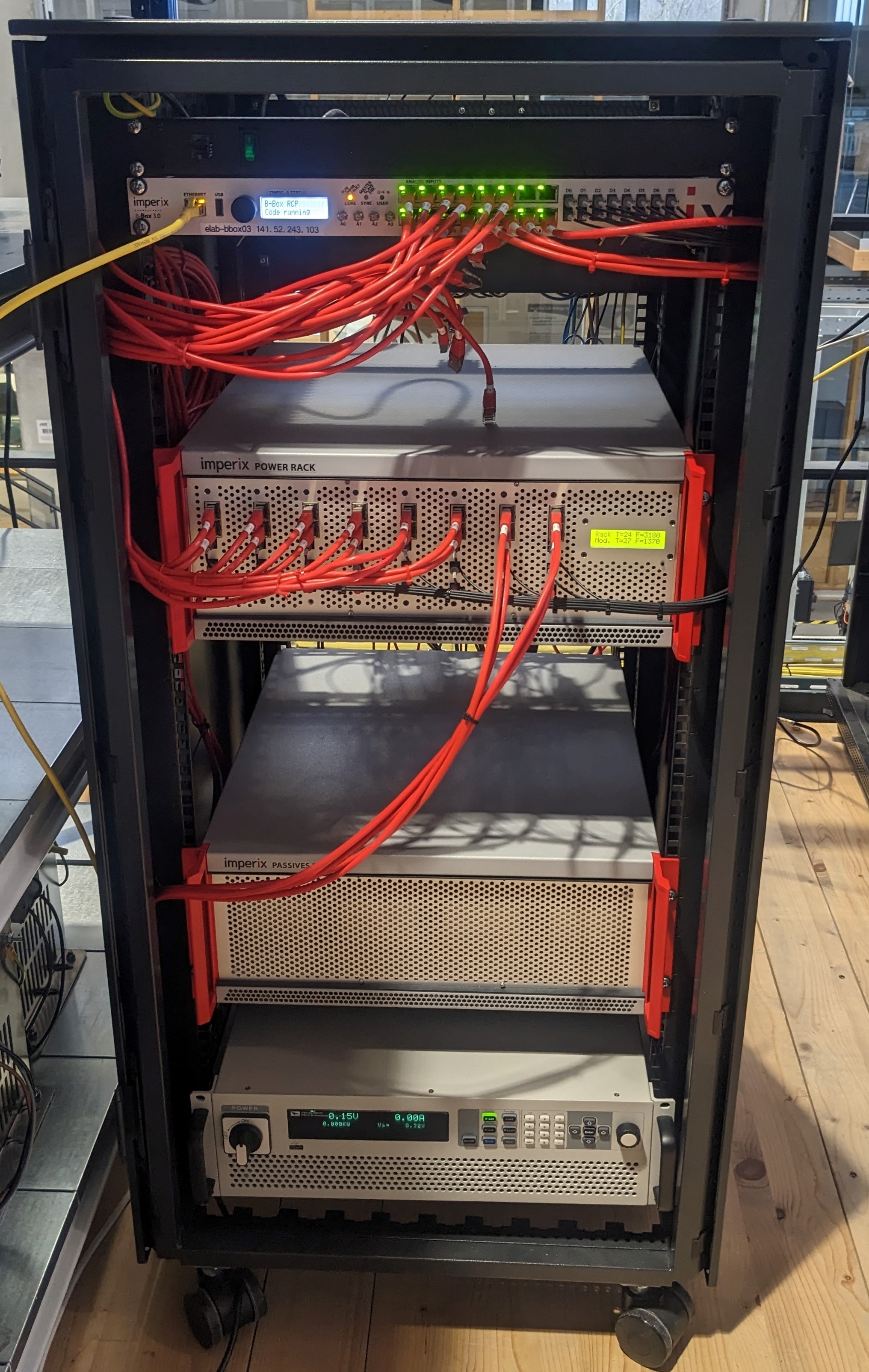}};
		\node [box,minimum width=2.2cm] at (4.45,3.0) (controller) {control unit};
		\node[left = 1.7cm of controller](lctrl){};
		\node [box,minimum width=2.2cm,below = 0.8cm of controller] (power) {\color{rev2}Power Rack: contains\\ \color{rev2} half-bridge modules\\ \color{rev2} for DC/DC and\\  \color{rev2}DC/AC converter};
		\node[left = 1.9cm of power](lpower){};
		\node [box,minimum width=2.2cm,below = 1.0cm of power,inner sep = 2pt] (passives) {\color{rev2} AC output filter \\ \color{rev2} and DC-leg inductance};
		\node[left = 2.5cm of passives](lpassives){};
		\node [box,minimum width=2.2cm,below = 0.2cm of passives] (dcsupply) {DC power supply};
		\node[left = 2cm of dcsupply](ldcsupply){};
		\draw[->,maigreen] (controller) -- (lctrl);
		\draw[->,maigreen] (power) -- (lpower);
		\draw[->,maigreen] (passives) -- (lpassives);
		\draw[->,maigreen] (dcsupply) -- (ldcsupply);
	\end{tikzpicture}
	\caption{Overview of a programmable DC/AC converter system for the hardware experiments.}
	\label{fig: setup}
\end{figure}
As shown in Fig.~\ref{fig: setup}, the hardware representation of the programmable DC/AC converter system consists of a DC power supply, DC/DC converter, a control unit, a DC/AC converter and an output filter. 
\subsubsection{Programmable DC/AC converter system}
\begin{figure}[ht]
	\centering
	\includegraphics[width=0.85\linewidth]{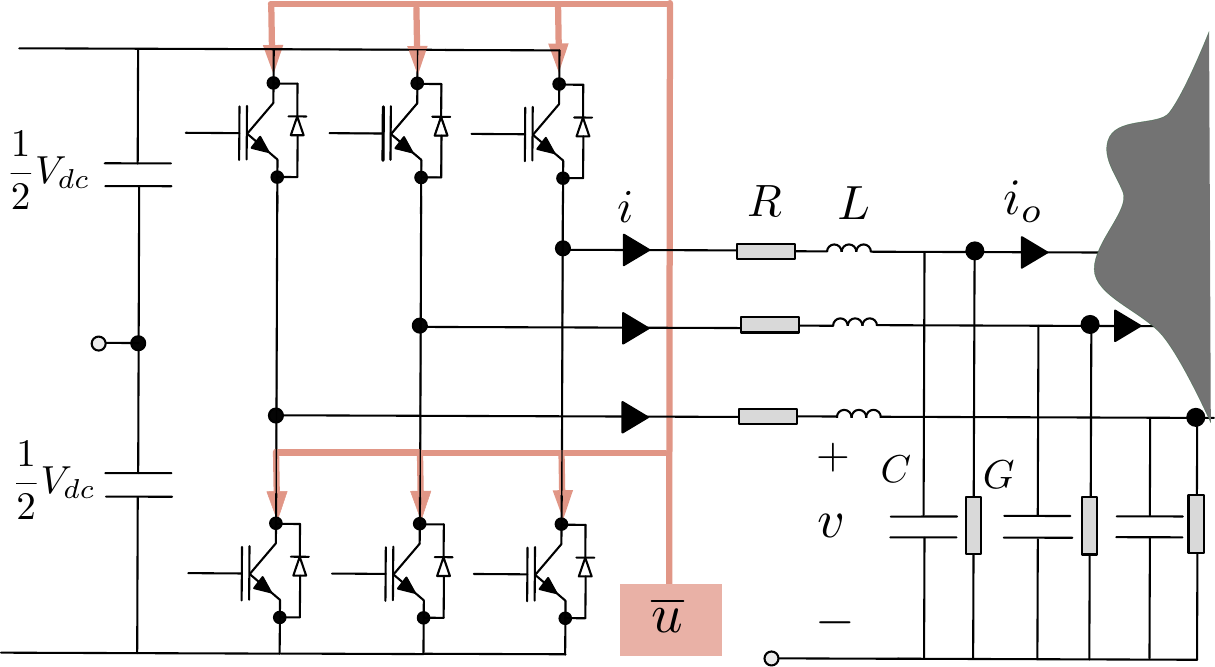}
	\caption{A three-phase DC/AC converter model under study. \textcolor{rev2}{Each phase-leg consists of a half-bridge module \cite{fn:PEB8038}}. The modulation input $\bar{u}$ is obtained either directly (direct control) or via well-known cascaded voltage and current control (indirect control), \textcolor{rev1}{for control diagrams and further details see Appendix \ref{ap:appendixA}.}}
	\label{fig: DCAC_converter}
\end{figure}

\begin{figure}[ht]
	\centering
	\includegraphics[width=0.85\linewidth]{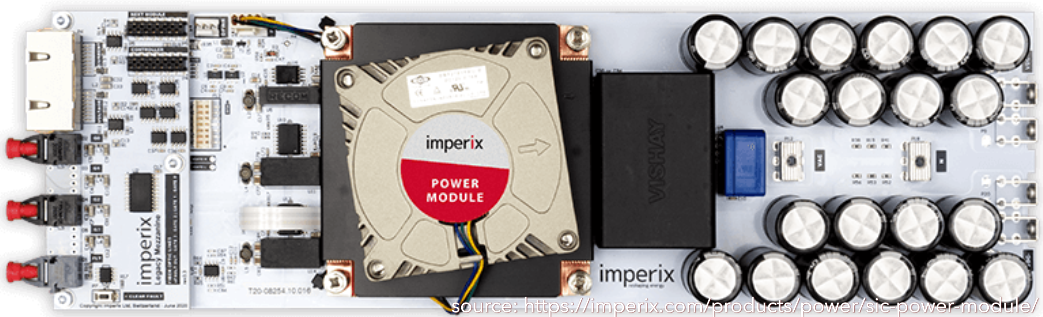}
	\caption{\textcolor{rev1}{Exemplary image of the half-bridge modules \cite{fn:PEB8038} contained in the power rack.}}
	\label{fig: halfbridge_leg}
\end{figure}

\begin{itemize}
	\item DC power supply: emulates the power source of the DC/AC converter system and has a rated power of $15\,\mathrm{kW}$.
	\item \textcolor{rev2}{DC/DC converter: consists of a boost converter which is build using a half-bridge module~\cite{fn:PEB8038} as well as an inductance of $L_b = 1.3\,\mathrm{mH}$. The system is controlled to regulate the DC bus voltage at the desired value.}
	\item \textcolor{rev2}{DC/AC converter and output filter: the two-level three-phase DC/AC converter system consists of SiC-MOSFET half-bridge modules, see Fig.~\ref{fig: halfbridge_leg}, including an embedded DC-bus of $500\,\mu \mathrm{F}$, as described in~\cite{fn:PEB8038}. To form the three-phase DC/AC converter, the half-bridge modules are connected on the DC-side to form a common DC-bus, which is supplied by the DC/DC converter with a constant voltage. The output filter consists of an inductance in series with a parasitic resistance for each of the phases. They are set in parallel with a capacitance to form a LC-filter~\cite{passiv-rack}.} The DC/AC converter system has a rated power of $P_{rated} = 15$ kW and a nominal voltage of $230\sqrt{2}\,\mathrm{V}$. The schematics are shown in Fig.~\ref{fig: DCAC_converter} 
	\item Control unit: the real-time control unit~\cite{B-Box} executes the user code and controls both the DC/DC and the DC/AC stage during the experiments. \textcolor{rev1}{The control cycle frequency of $20\,\mathrm{kHz}$ is well above the nominal grid frequency of $50\,\mathrm{Hz}$ such that the control latency has only negligible effect on the presented results.}
    \item \textcolor{rev1}{Measurement sensors: To obtain the necessary measurements for the control of the converters and presentation of the results, the following sensors are used:
        \begin{itemize}
            \item Current sensors \cite{currentsensor}: the measurement range of $\pm 50\,\mathrm{A}$ is well suited for this application and the bandwidth of $200\,\mathrm{kHz}$ is large enough to cover the relevant frequency range, considering a switching frequency of $20\,\mathrm{kHz}$ and nominal grid frequency of $50\,\mathrm{Hz}$. Further, the typical sensitivity error of the sensors is $\pm 0.4\,\%$ which is sufficient for the control usecase. The input-referred noise is $0.05\,\mathrm{A}$, which is several orders of magnitude smaller than the operating point during the experiments and therefore has negligible effect on the presented results.
            \item Voltage sensors \cite{voltagesensor}: the measurement range of $\pm 800\,\mathrm{V}$ and the bandwidth of $100\,\mathrm{kHz}$ is sufficient to capture the relevant dynamics, considering the switching frequency of $20\,\mathrm{kHz}$ and nominal grid frequency of $50\,\mathrm{Hz}$. Further, the typical sensitivity error of the sensors is $\pm 0.35\,\%$ which is adequate for the control usecase. The input-referred noise is $1.4\,\mathrm{V}$, which is sufficiently small compared to the operating voltage.
        \end{itemize}
        Since each phase voltages and currents are obtained by individual sensors, the values for the sensitivity and offset differ slightly even after calibration.
        Such sensor-specific errors can lead to an imbalance in the measured phase currents and voltages, which generates inter-harmonic components in derived values such as the active power.}
\end{itemize}
The parameter values of the programmable DC/AC converter are summarized in Table \ref{tab: dcac_param}.

\subsubsection{Resistive load}
\begin{figure}[h!]
	\centering
	\includegraphics[width=0.85\linewidth]{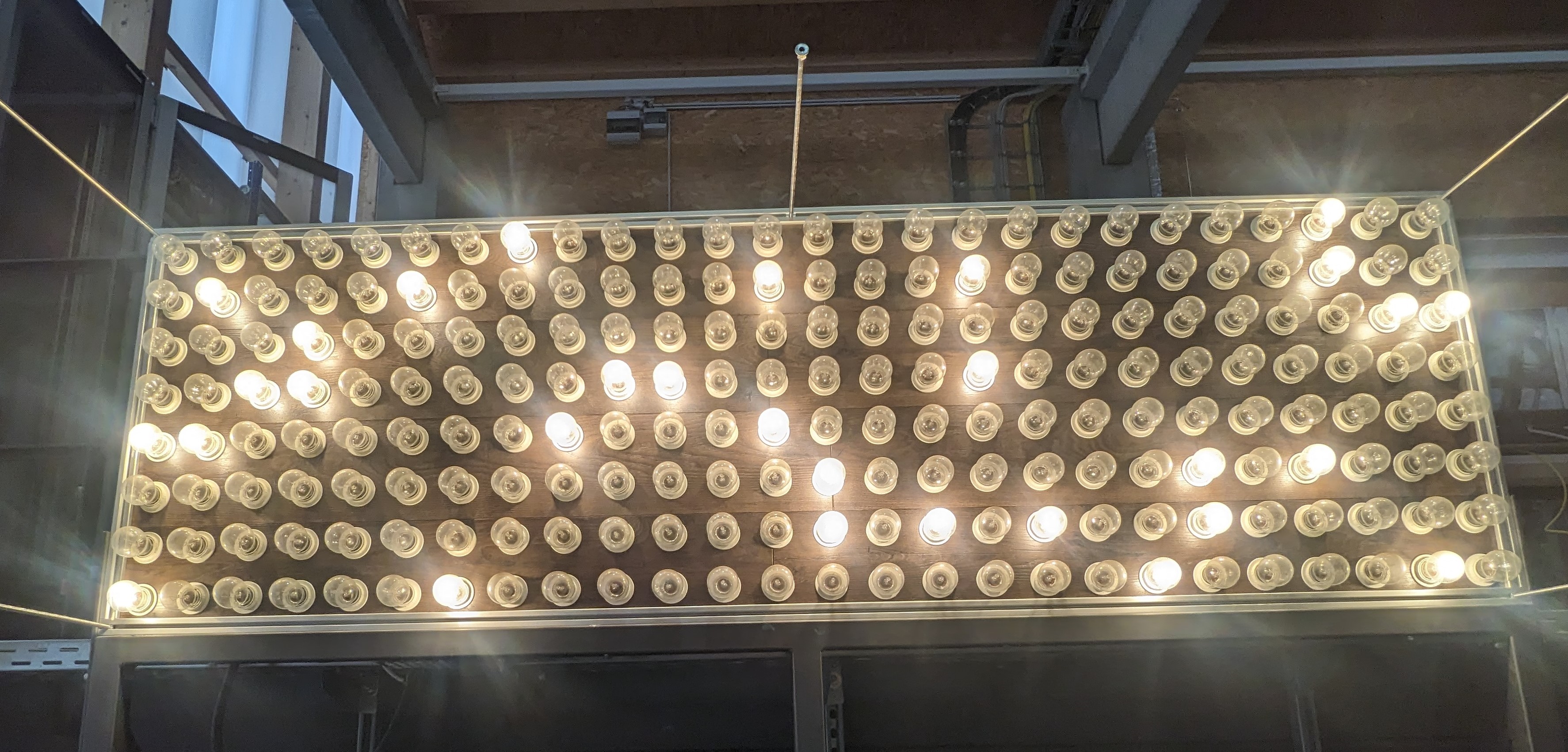}
	\caption{Resistive load represented by incandescent light bulbs.}
	\label{fig: light_wall}
\end{figure}

We consider a light-wall to represent a resistive load, see Fig.~\ref{fig: light_wall}. Each of the light bulbs has a power consumption of $100\,\mathrm{W}$ at nominal voltage of $230\sqrt{2}\,\mathrm{V}$ and can be individually controlled via a programmable logic controller to set the desired power consumption.

\subsubsection{Transmission line replica}
\begin{figure}[h!]
	\centering
	\includegraphics[width=0.6\linewidth]{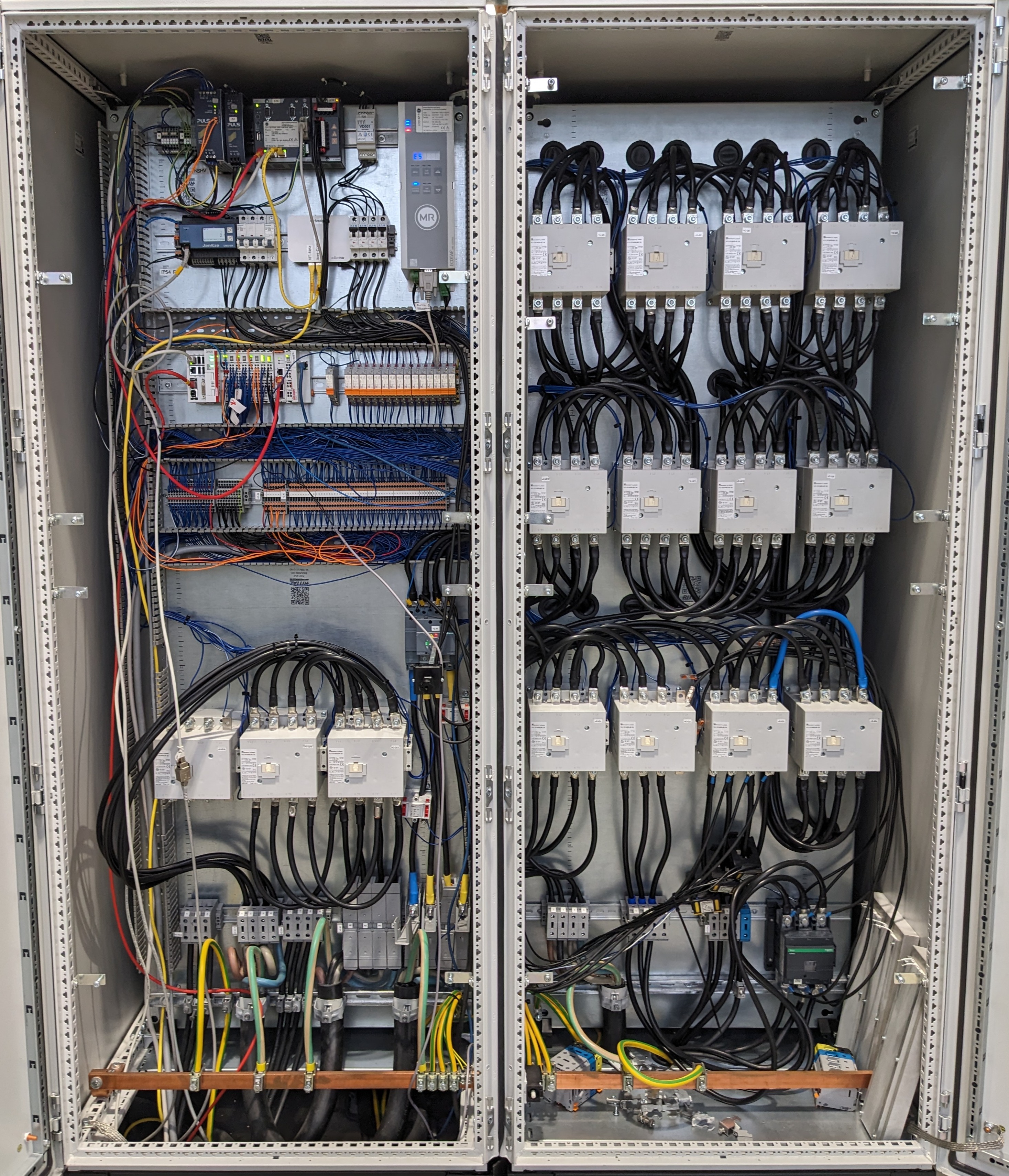}
	\caption{\textcolor{rev1}{Transmission line replica in our experimental setup.}}
	\label{fig:transmissionline}
\end{figure}

The physical transmission line properties are emulated using resistive and inductive type line replicas. \textcolor{rev1}{Discrete hardware elements are connected in series to provide the desired physical quantities in order to replicate the effect of a given length and type of transmission line~\cite{wiegel2022smart}, see Fig.~\ref{fig:transmissionline}.}
The parameter values of the transmission line of our experimental setup are given in Table~\ref{tab: dcac_param}.

\begin{table}[h!]
	\caption{Technical details of the hardware setup}
	\centering 
	DC/AC converter and its control
	\begin{center}
			\begin{tabular}{lllc}
				\hline
				Symbol & Definition  & Value (S.I.)\\
				\hline
				$V^{*}$ & AC voltage amplitude & $230 \sqrt{2}$\\
				$C$ & AC filter capacitance & $1 \cdot 10^{-5}$\\
				$L$ & AC {filter} inductance & $2.36\cdot 10^{-3}$\\
				$R$ & AC {filter} resistance & $1 \cdot 10^{-3}$\\
				$G$ & load resistance & $58.77$\\
                $V_{dc}$ & DC-link voltage &  $750$\\
                \color{rev1}$f_{sw}$ & \color{rev1}switching frequency &  \color{rev1}$20\,\mathrm{kHz}$\\
				\hline
				$A$ & modulation amplitude  & $0.8132$\\		
				\hline
			\end{tabular}
	\end{center}
	\label{tab: dcac_param}
	\centering 
	Angular droop control
	\begin{center}
			\begin{tabular}{lllc}
				\hline
				Symbol & Definition  & Value (S.I.)\\
				\hline
				$P^*$ & nominal active power& $2880$\\
				$\omega^*$ & nominal angular frequency& $2\pi 50$\\
				\hline
				$\alpha$ & input effort gain& $2000$\\
				$\gamma$ & steady state gain & $5\cdot 10^{4}$\\
				\hline
				$k_{VP}$ & P-voltage gain  & $0.05$\\
				$k_{VI}$ &  I-voltage gain & $0.4$\\
				$k_{IP}$ & P-current gain & $10$\\
				$k_{II}$ & I-current gain & $240$\\
				\hline
			\end{tabular}
	\end{center}
	\label{tab-angular scenarioI}
	Parameters values of the transmission line
		\begin{center}
						\begin{tabular}{lllc}
								\hline
								Symbol & Definition  & Value in S.I.\\
								\hline
								$R_l$ & line resistance & $20\cdot 10^{-3}$\\
								\hline
								$L_l$ & line inductance & $700\cdot 10^{-6}$\\
								\hline
						\end{tabular}
			\end{center}
		\label{tab: trans-line}
		\vspace{-0.3cm}
\end{table}
\color{black}

\section{Scenario I: Single converter to load}
\label{sec:scenario-I}
Angular droop control operates in a different domain than conventional frequency droop by linking active power to angle rather than frequency. Although angle and frequency are related through integration, this distinction poses challenges for the real-world implementation.
\subsection{Scenario Goals and Description}
\begin{figure}[ht!]
    \centering
    \includegraphics[width=.7\linewidth]{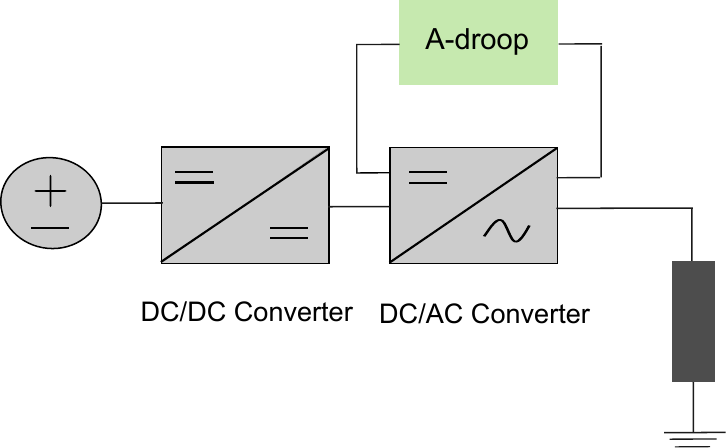}
    \caption{Schematic representation of Scenario I consisting of a  balanced three-phase DC/AC converter in closed-loop with angular droop control and connected to a resistive load.}
    \label{fig: scenarioI}
\end{figure}

To experimentally verify the properties of the angular droop hardware implementation under real-world conditions, we start with Scenario I: a single DC/AC converter system in closed-loop with the angular droop control is connected to a resistive load, as shown in Fig.~\ref{fig: scenarioI}.

With Scenario I we show:
	\begin{itemize}
		\item black start capabilities, namely the ability to form sinusoidal wave after a major event, e.g., a blackout.
		\item the capability to withstand load disturbance, e.g., upon a sudden increase/decrease in the load power.
		\item influence of direct and indirect implementation schemes on the angular droop behavior and drawing a comparison between them. The modulation input $\bar{u}$ of the DC/AC converter is obtained either directly (direct control) or via well-known cascaded voltage and current control (indirect control), details see in Appendix \ref{ap:appendixA}.
	\end{itemize} 
    The single-converter test case is crucial for assessing the resilience of the closed-loop converter system against disturbances of varying magnitudes. The ability to withstand small to medium disturbances is particularly relevant in conditions close to nominal operation, while grid-forming capabilities become essential following major disruptions such as blackouts.
    Comparing direct and indirect implementations shows the compatibility of angular droop with well established control architectures, while preserving its properties.
\subsection{Challenge: Discretization of the angle dynamics}
\paragraph{Analysis}
The angular droop control described in Section~\ref{sec:Understand-adroop-ctrl}, also see Appendix~\ref{ap:appendixA}, needs to conform with the restrictions posed by real hardware, in particular its implementation in discrete-time. For this, we discretize the closed-loop angle dynamics in~\eqref{eq: cl-dyn} using the forward Euler method as follows,
\begin{align}
\label{eq: disc-angle-dyn}
\theta(s+1) &= \theta(s) + T_s \,  u_d(s) +\omega^*\mathds{1}_n \,,\\ \nonumber
u_d(s)&= -\frac{1}{2}R^{-1}\, (\Gamma\, (\theta(s)-\theta^*(s))+ P(\theta(s))-P^*) \,,
\end{align}
where $s\in\mathbb{Z}$ is the time step, $T_s>0$ is the sampling period and  $\omega^*>0$ is the nominal angular frequency.
Next, we define the angle error coordinate $\Delta \theta(s) = \theta(s) - \theta^*(s)$ where $\theta^*\in\real^n$ denotes the nominal angle satisfying
\begin{align}
\label{eq: ss-angle-dyn}
\theta^*(s+1) = \theta^*(s) + T_s \,\omega^*.
\end{align}
From \eqref{eq: disc-angle-dyn}, the discrete-time angle error dynamics are given by,
\begin{align}
\label{eq: error-dyn}
\Delta \theta (s+1) &= \Delta \theta (s) - \frac{1}{2} T_s \, R^{-1} \, (\Gamma \Delta \theta(s)+ P(\theta(s)) - P^*).
\end{align}
Observe that in~\eqref{eq: ss-angle-dyn}, the nominal steady state angle vector $\theta^*$ grows infinitely. This causes a loss of precision for the stored variable $\theta(s)$ due to the limit of available bytes for single precision.
\paragraph{Proposed solution}
The previous observation motivates the following solution. We aim to find a mapping of the angles $\theta^*$ from $\real^n$ to the $n-$th dimensional torus $\mathbb{T}^n$.
Note that the sine function appearing in the implementation of the modulation signal $\overline u$ both for the direct~\eqref{eq: angular droop-impl} and indirect~\eqref{eq: vol_ref} schemes is $2\pi$ periodic in the angle $\theta(s)$. We proceed by limiting the values of the nominal angle $\theta^*(s)$ as follows,
\begin{align}
\label{eq: mod-2pi}
\theta^*(s+1) = \theta^*(s)+T_s\,\omega^* (\mathrm{mod\,2\pi}),
\end{align}
which yields the following absolute angles,
\begin{align}
\theta (s+1) = \theta^*(s+1) + \Delta \theta (s+1).
\end{align}
Here, $\Delta \theta (s+1)$ is given by~\eqref{eq: error-dyn}.
Therefore the angles $\theta(s), \, s\in\mathbb{Z}$ remains within feasible numerical bounds.
Our proposed solution optimizes the space complexity of hardware implementation while ensuring ease of understanding and practicality.
\subsection{Experimental Results}
\label{sec:experimentalresults}
    In the following, we present our main observations from the experiments of Scenario I. All unis are in S.I.
	\subsubsection{Black start capabilities}
\begin{figure}
		\centering
		\includegraphics[width=0.85\linewidth]{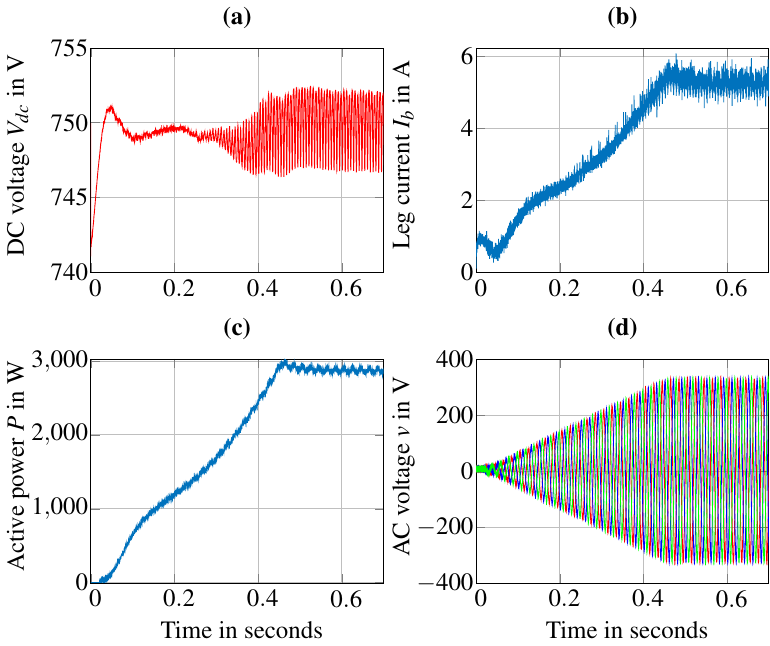}
		\caption{Averaged DC voltage $V_{dc}$ in (a), DC/DC leg inductance current $I_b$ in (b), active power $P$ in (c) and AC voltage $v$ in (d) in the black start experiment.}
		\label{fig:dc-ac-direct-black}
	\end{figure}

	From Fig.~\ref{fig:dc-ac-direct-black}, we observe that the DC capacitor voltage $V_{dc}$ reaches its nominal value $V^{nom}_{dc}= 750$ within $0.4\,\mathrm{s}$ following initial transients. This corresponds to the leg inductance current converging to the steady state value $I^{s}_b\approx 5$.
	Following the transients, the visible voltage ripple is due to the sizing of the DC-link capacitors and remains within an acceptable range for the given resistive load. On the AC side, the active power converges to its nominal value $P^*=2880$ which corresponds to a sinusoidal balanced three-phase signal of the output capacitor voltage $v$ with nominal amplitude $V^d= 230\sqrt{2}$. In essence, Figure~\ref{fig:f-thetangular blackstart} shows that both the angle and frequency error converge to zero, i.e., the AC frequency and the phase angle of the modulation signal are at their nominal values, namely $\omega^* = 2\pi\,50$ and $\theta^*(t)=2\pi\,50  \,\mathrm{t}$, respectively.
	Thus, our results demonstrate that, even if we start from initial operating conditions far away from nominal operation, resulting from large disturbances, e.g., a black start, the angular droop control is able to form sine waves rotating at a nominal frequency with desired angle and amplitude and is therefore grid-forming.
	\subsubsection{Step in the load power}
    \begin{figure}
		\centering
		\includegraphics[width=0.85\linewidth]{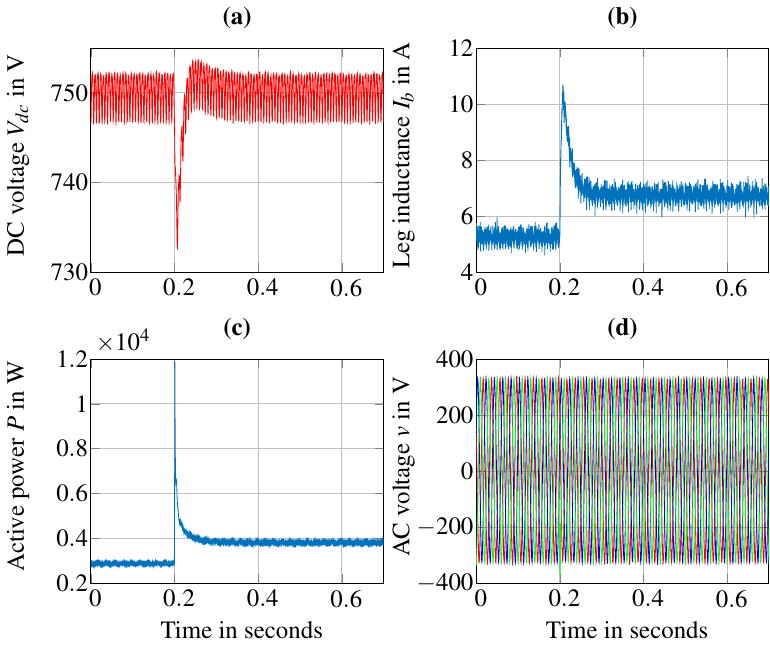}
		\caption{Averaged DC voltage $V_{dc}$ in (a), DC/DC leg inductance current $I_b$ in (b), active power $P$ in (c) and AC voltage $v$ in (d) in the load step experiment.}
		\label{fig:dc-ac-direct-loadstep}
	\end{figure}
\begin{figure}
		\centering
		\includegraphics[width=0.85\linewidth]{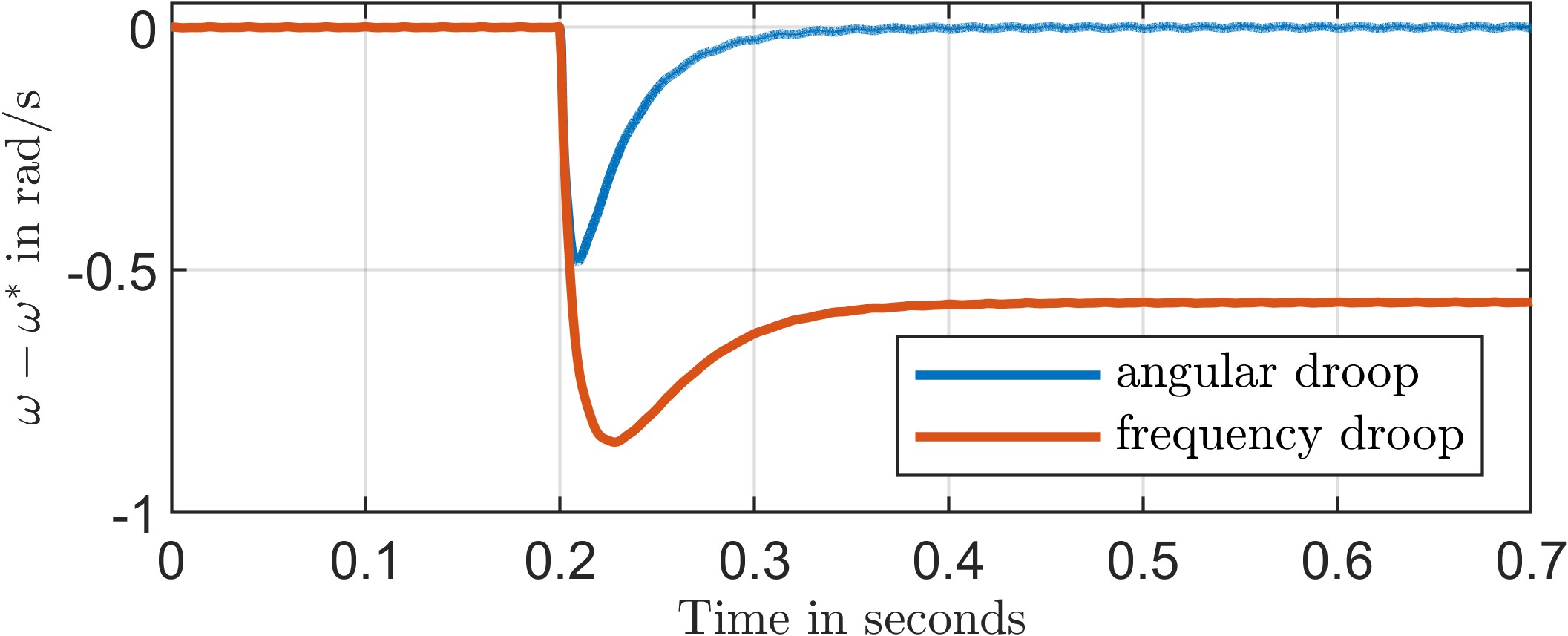}
		\caption{\textcolor{rev2}{Comparison of angular droop and frequency droop for the load step experiment. Angular droop control uses $\alpha = 2000,\,\gamma = 5\cdot 10^4$ and frequency control a $5\%$ droop constant.}}
		\label{fig:angular_vs_frequency_droop_loadstep}
	\end{figure}

    Proceeding from the steady state after the black start, we study a load step change at $t= 0.2\,\mathrm{s}$. The active power overshoots to approximately $1.2\cdot 10^4$ and settles to a new steady state $P^{s} \approx 3800$, see Fig.~\ref{fig:dc-ac-direct-loadstep}. The overshoot is caused by large rush currents accompanying the change in load power. The DC voltage $V_{dc}$ returns to its nominal value after $0.1\,\mathrm{s}$ due to the integral control action of the DC voltage controller, whereby the DC/DC leg inductance current reaches a newly induced steady state $I_b^{s}\approx 7$ following the load step. The three-phase voltage amplitude drops slightly during the event and recovers to its nominal value at steady state. 
     \textcolor{rev2}{Figure~\ref{fig: f-thetangular load-step} shows a drop in the AC frequency during the load power change, however the frequency returns to its nominal steady state despite the increase in load power.
     This zero steady state frequency error is an advantageous, intrinsic property of angular droop control as previously discussed in Sec.~\ref{sec:angle_frequency_droop_comp} and makes the deployment of secondary control to restore the frequency to its nominal value obsolete. Whereas, in a frequency droop controlled setup such a step in the load power causes permanent frequency deviation, unless the power reference is adapted by the secondary control layer.
     This is shown in Fig.~\ref{fig:angular_vs_frequency_droop_loadstep}, where angular droop is compared to frequency droop upon a load step. Note that the steady state deviation for the frequency droop depends on the chosen droop constant.}

\begin{figure}
		\centering
		\includegraphics[width=0.85\linewidth]{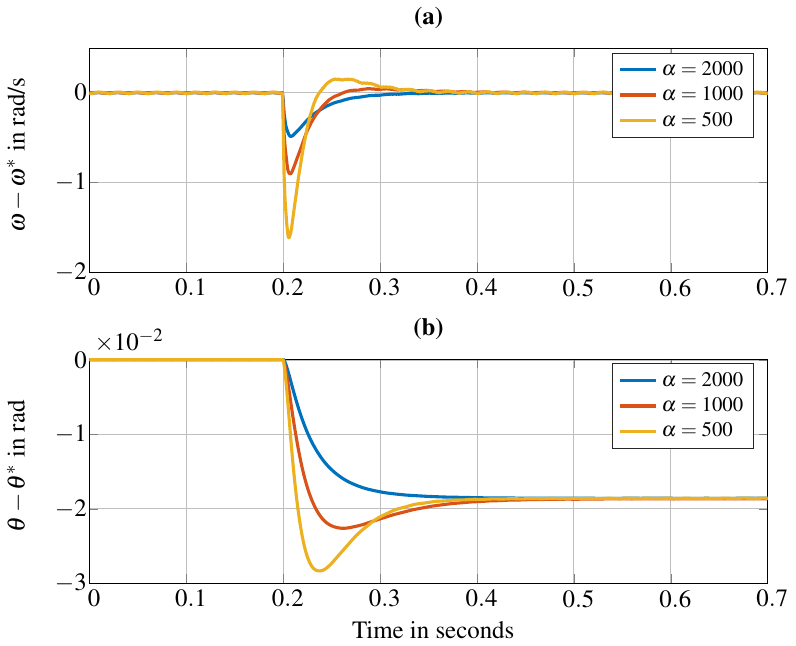}
		\caption{Frequency (a) and angle errors (b) with $\gamma=5 \cdot 10^4$ and different values for $\alpha\in~\{ 500,\, 1000,\,  2000\}$ in the load step experiment.}
		\label{fig: alpha}
	\end{figure}

\begin{figure}
		\centering
		\includegraphics[width=0.85\linewidth]{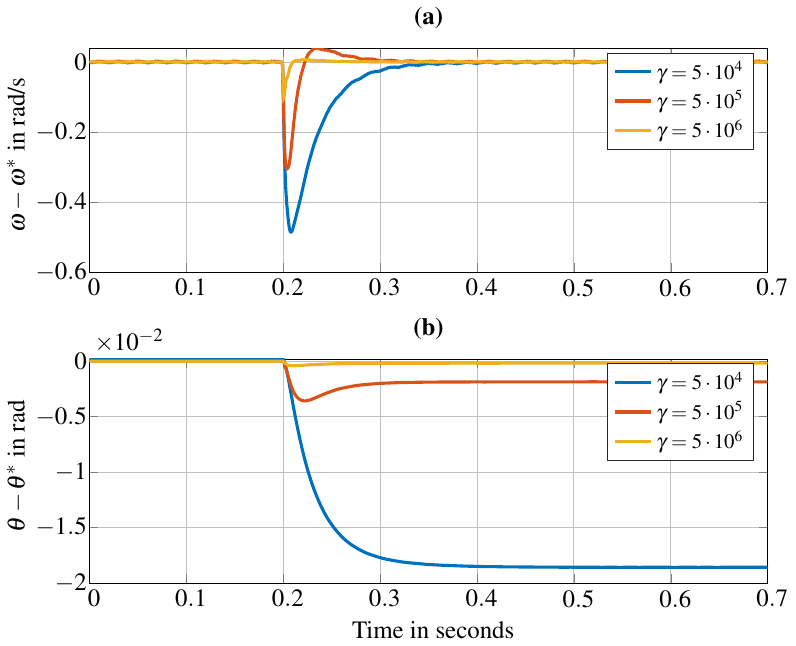}
		\caption{Frequency (a) and angle errors (b) with $\alpha=2000$ and different values for $\gamma \in \{ 5\cdot 10^4,\, 5\cdot 10^5,\, 5\cdot 10^6\}$ in the load step experiment.}
		\label{fig: gamma}
	\end{figure}

	In the following, we determine a suitable tuning of the angular droop control by studying the influence of the control gains $\alpha,\gamma>0$ on the frequency and angle of the modulation signal following a step in the load at $t= 0.2\,$s. 
	As depicted in Fig.~\ref{fig: alpha}, an increase in $\alpha$ leads to a higher penalty on the angle and power deviations from their steady state. This translates into a larger rate of change of frequency and smaller nadir. This is in accordance with the observation that, the choice of the gain $\alpha$ affects the transient behavior of the AC frequency $\omega$ and angle $\theta$ in~\eqref{eq: rocof}. To avoid an overshoot and limit the frequency deviation to an acceptable deviation of e.g., $0.8$ Hz~\cite{freq_reg}, we fix the value $\alpha = 2000$.
	Figure~\ref{fig: gamma} shows that, for decreasing values of $\gamma$ the power-to-angle droop behavior is more pronounced resulting in larger steady state angle deviations, see~\eqref{eq: droop-bhv}. The induced steady state angle $\theta^{s}$ is given by
	\begin{align}
		\label{eq: ind-ss-angle}
		\theta^{s} = \theta^*+\frac{1}{\gamma} (P^*-P^{s}),
	\end{align}
	where $P^{s}$ is the load power at induced steady state following a step in the load. For different $\gamma$ values the rate of change of the angles during the transient leads to frequency behavior described by~\eqref{eq: rocof}. This empirically confirms that $\gamma$ affects both the transients of the AC frequency $\omega$ and steady state behavior of the angle $\theta$.
	Since the choice of $\gamma\geq 5 \cdot 10^5$ leads to relatively small power-to-angle droop behavior, we select $\gamma=5 \cdot 10^4$.
	\subsubsection{Comparison between direct and indirect control}
\begin{figure}
		\centering
		\includegraphics[width= 0.85\linewidth]{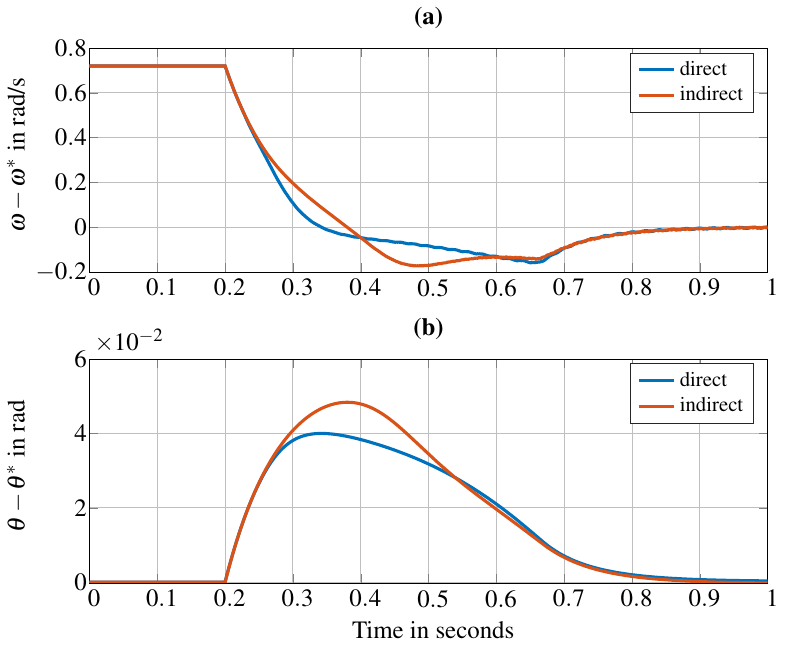}
		\caption{Frequency (a) and angle errors (b) for the black start experiment.}
		\label{fig:f-thetangular blackstart}
	\end{figure}

\begin{figure}
		\centering
		\includegraphics[width= 0.85\linewidth]{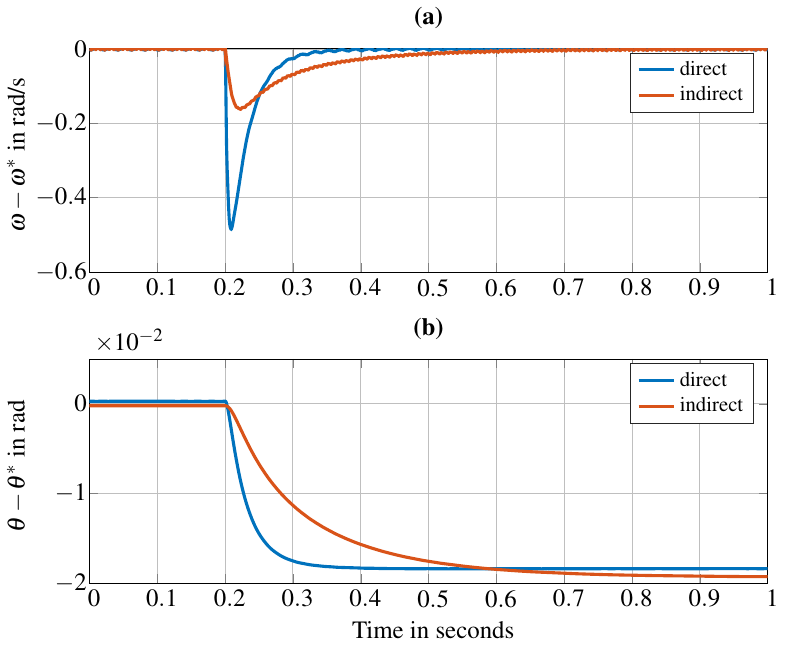}
		\caption{Frequency (a) and angle errors (b) in the load step experiment.}
		\label{fig: f-thetangular load-step}
	\end{figure}

\begin{figure}
		\centering
		\includegraphics[width =0.85\linewidth]{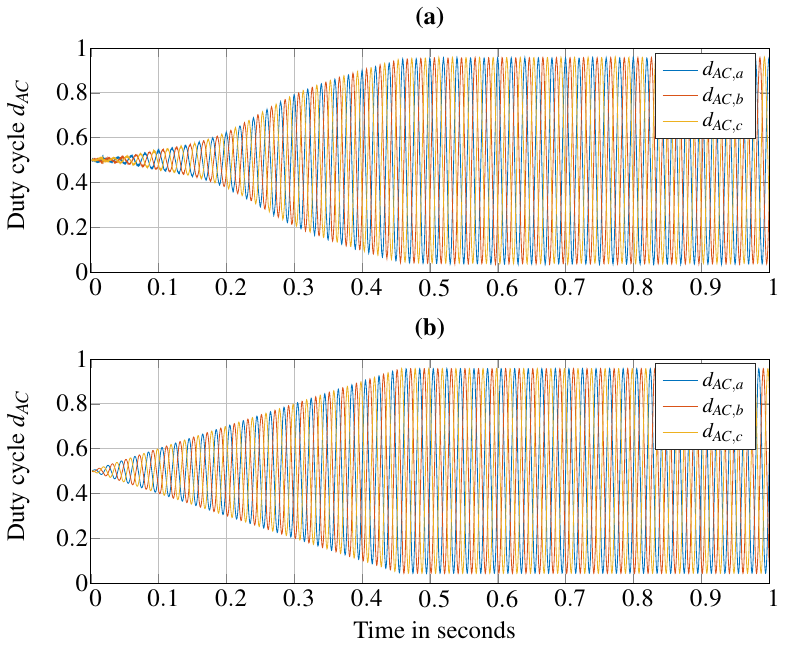}
		\caption{Duty cycles $d_{AC}$ of the indirect (a) and direct control (b) schemes in the black start experiment.}
		\label{fig: duty-cycles}
	\end{figure}

    For the hardware implementation, the angle dynamics resulting from the angular droop control can be translated to a modulation input $\bar{u}$ either directly (direct control) or via well-known cascaded voltage and current control (indirect control). For more details see Appendix~\ref{ap:appendixA}.
	For the control gains $\alpha=2000$ and $\gamma = 5\cdot 10^4$, Fig.~\ref{fig:f-thetangular blackstart} and~\ref{fig: f-thetangular load-step}
	compare the frequency and angle errors resulting from the direct and indirect implementation for the black start and the load step experiment, respectively.
    \textcolor{rev1}{In Table~\ref{tab:comparisonmetrics_direct_indirect} numerical values for the comparison are presented.}
    Using the indirect implementation, the frequency is restored to its nominal value within approximately $0.61\,\mathrm{s}$ and the angles converge to an induced steady state angle $\theta^{s}$ as in Eq.~\eqref{eq: ind-ss-angle}.
	In fact, the major difference between direct and indirect control is in the way the control law~\eqref{eq: angular droop-impl} relates to the modulation signal representing the main input to the DC/AC converter. In the direct implementation, we assign a sinusoidal wave whose angle is directly determined by the angular droop control. The indirect implementation consists of cascaded voltage and current control loops relying on tracking a given AC voltage reference, whose angle is described by the angular droop control. This results in different modulation signals $\bar{u}$, where the control effort solely dependent on the gains $\alpha$ and $\gamma$ in the direct scheme. For the indirect scheme, the control effort is also dependent on the choice of the current and voltage control loops, see Table~\ref{tab-angular scenarioI}. This is visible from the duty cycles $d_{AC}$ in Fig.~\ref{fig: duty-cycles}, defined by~\eqref{AC-duty-cycle}. The duty cycle increases linearly in the direct whereas sub-linearly in the indirect implementation scheme. For our particular control gain choice, this results in a slower convergence rate to a steady state compared to the direct scheme for the load step experiment, \textcolor{rev1}{see Table~\ref{tab:comparisonmetrics_direct_indirect}}. 
	Due to its simpler and more intuitive tuning, we adopt the direct implementation of the angular droop control with $\alpha= 2000$ and $\gamma = 5 \cdot 10^4$ in the remainder of our hardware experiments.
\begin{table}[h!]
	\caption{\textcolor{rev1}{Comparison between direct and indirect control. RMS denotes the root mean square, nadir the smallest value, max. the maximum value and $T_\mathrm{set}$ the settling time to $\pm0.02 \mathrm{Hz}$ frequency deviation following the event.}}
	\centering 
        Black start experiment\\
			\begin{tabular}{l c c c c c}
				\hline
				control  & nadir $\Delta \omega$  & RMS $\Delta \omega$ & max. $|\Delta \theta|$ & RMS $\Delta \theta$&$T_\mathrm{set}$\\
				\hline
                direct  & -0.159  & 0.165  & 0.04  &  0.024 & 0.61\\
                indirect  & -0.173  & 0.181  & 0.48  &  0.027 & 0.61 \\
				\hline
			\end{tabular}
            
        Load step experiment\\
			\begin{tabular}{l c c c c c}
				\hline
				control  & nadir $\Delta \omega$  & RMS $\Delta \omega$ & max. $|\Delta \theta|$ & RMS $\Delta \theta$ & $T_\mathrm{set}$\\
				\hline
                direct  & -0.485  & 0.082  & 0.018  & 0.0178  & 0.11\\
                indirect  &  -0.162 & 0.045  & 0.019  &  0.0170 & 0.24\\
				\hline
			\end{tabular}
	\label{tab:comparisonmetrics_direct_indirect}
\end{table}
\section{Scenario II: Two identical converters to a common load}
\label{sec:scenario-II}
The use of angles instead of frequency has further implications for the real-world application of angular droop in grids with multiple converters. Unlike nominal frequency, which is constant, the nominal angle changes over time. This must be considered for the parallel operation of distributed DC/AC converters controlled by angular droop.
\subsection{Scenario description}
\begin{figure}[ht!]
	\centering
	\includegraphics[width=\linewidth]{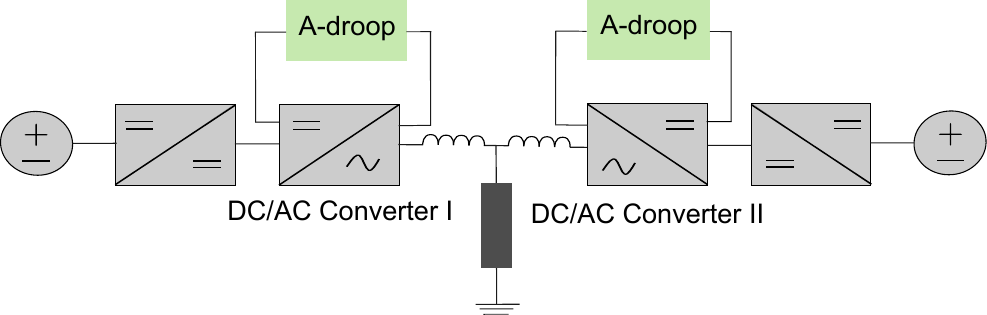}
	\caption{Schematic representation of Scenario II consisting in two identical converter systems, each in closed-loop with angular droop control, connected to a common resistive load via highly inductive transmission lines.}
	\label{fig: scenarioII}
\end{figure}

Scenario II consists of two identical DC/AC converter systems supplied by two independent DC sources behind DC/DC converters and connected to a common resistive load as shown in Fig.~\ref{fig: scenarioII}.
Table~\ref{tab: trans-line} summarizes the parameter values of the transmission line replicas.
To satisfy the modeling assumption of highly inductive transmission lines, we set the ratio $X/R_\ell\approx 11$ with the reactance $X = \omega^* L_\ell$, see also~\cite{kolluri_2017}.
In this scenario, we verify:
	\begin{itemize}
		\item frequency synchronization capabilities of the angular droop control.
		\item power-sharing capabilities in dependence of the angular droop control gains.
	\end{itemize}

 We hereby underscore the relevance of the two-converter test case as a toy example that provides a solid foundation for generalizations towards $n-$converter system with $n>2$ for the following reasons. First, frequency synchronization and power-sharing are fundamental properties that are shared by a network of converter of any size $n>1$. Second, any analysis comprising $n-$converters with $n>2$ can be divided into a pairwise study of two converter $i$ and $j$ and therefore reduces to a two-converter setup. Third, the same implementation challenges, in particular, {\em clock drifts} are encountered in any extension to a network.
\subsection{Challenge: Clock drift in angular droop controlled grids}
\label{subsec: Challenge_2}
\paragraph{Analysis}
Angular droop control is susceptible to clock drifts. This is well-documented in the power system literature~\cite{kolluri_2017, 7781662} and can be explained as follows.
Let $t_k>0$ denote the local time at the $k$-th DC/AC converter with respect to a global reference time $t$ as follows,
\begin{align}
t_k := (1+\epsilon_k)t,
\end{align}
where $\epsilon_k>0$ is the time-invariant drift of the local clock with respect to the reference clock. This drift arises in the absence of a master clock. Starting from a nominal global frequency $\dot\theta^{s}_k = \omega^*$, the angle obtained from the integration (assuming zero initial conditions) is affected by the clock drift because
\begin{align}
\theta^{s}_k(t) = \omega^*\, t_k = \omega^* \, (1+\epsilon_k)\, t = \omega_k\, t,
\end{align}
where $\omega_k := \omega^* \, (1+\epsilon_k)$ is local frequency at the $k-$th converter. Under the assumption of a highly inductive, Kron-reduced~\cite{6316101} power network, the active power at the output of the $k$-th converter at steady state is given by~\cite{kundur1994power}
\begin{align}
P^{s}_{k} &= \frac{V_k\,V_j}{X_{kj}} \sin(\theta^s_k-\theta^s_j) = \frac{V_k\,V_j}{X_{kj}} \sin((\omega_k-\omega_j)t)\\
& = \frac{V_k\,V_j}{X_{kj}} \sin(\omega^*\, t\, (\epsilon_k-\epsilon_j)),
\end{align}
where $\theta^s_j(t):= \omega_j\, t$ is the steady state angle, $\omega_j$ is the local frequency at the $j-$th converter and $X_{jk}$ is the reactance between converters $k$ and $j$ following Kron-reduction.
It can be deduced that when local clock drifts are not compensated for, the injected active power $P^{s}_{k}$ drifts apart from its nominal value.
Since the angular droop control law involves an integration, the closed-loop dynamics~\eqref{eq: cl-dyn} are not robust to local clock drifts~\cite{7781662} and suitable solution needs to be developed for the hardware implementation. 
\paragraph{Proposed solution}
In our lab experiments, we utilize the distribution of a common high-frequency clock~\cite{imperixsynch} (or master clock) across the entire control network through a direct optical fiber connection between the controllers of Converter I and II~\cite{clocksynch}. Therefore, the distributed devices belong to same clock domain, which eliminates the clock drift for the integration actions.
Our proposed solution enables a simple yet resource-friendly realization of angular droop control for testing multi-converter cases and provides a synchronization accuracy of $\pm 2\mathrm{ns}$ \cite{imperixsynch}, surpassing the clock accuracy of the global positioning system (GPS), which is in the range of $\pm 10\,\mathrm{ns}$ to $\pm 100\,\mathrm{ns}$~\cite{clocksynch}.
\subsection{Experimental results}
\subsubsection{Frequency synchronization}
\begin{figure}[ht!]
	\centering
	\includegraphics[width=\linewidth]{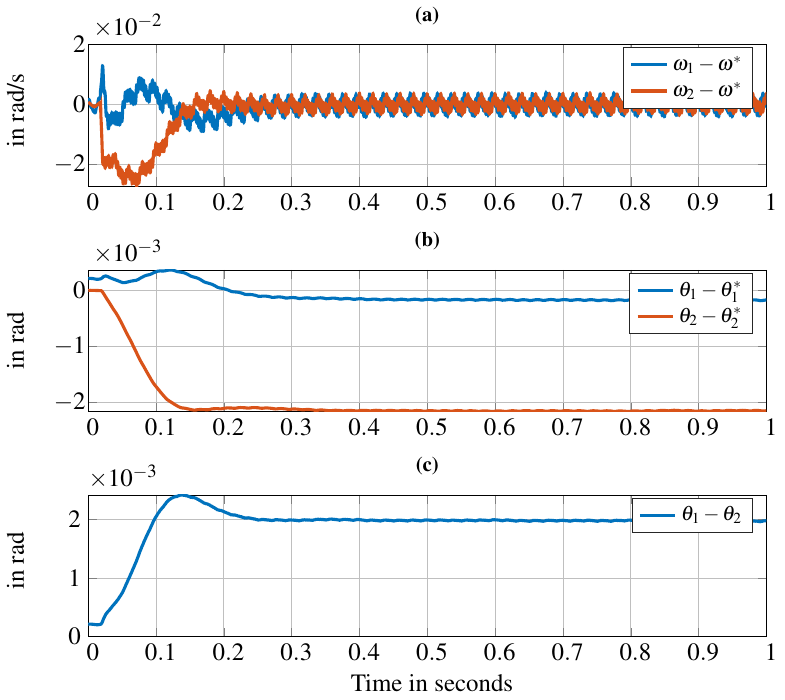}
	\caption{Frequency (a)  and angle error (b) for $k = 1,2$ and angle differences (c), following the connection of Converter II to Converter I at $t = 0$, both in  closed-loop with the direct angular droop control for $\gamma_k=5 \cdot 10^4$ and $\alpha_k=2000$ for $k = 1,2$. Here $\theta^*_1(t)=\omega^*t$ for $t<0$ and $\theta^*_2(t)=\omega^*t+\theta_2^*(0)$ for $t\geq 0$, where $\theta_1(0)$ is the modulation angle of Converter I at the time of interconnection.}
	\label{fig:TCL_sync}
\end{figure}

\begin{figure}[ht!]
	\centering
	\includegraphics[width=.7\linewidth, trim = {0 0.5 0 0.01cm}, clip]{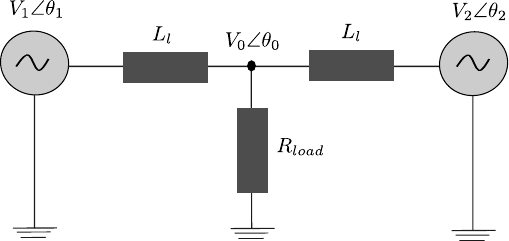}
	\caption{Representation of Scenario II as two voltage sources with switching voltage amplitude $V_k$ and modulation angle $\theta_k$ for $k=1,2$ supplying a common resistive load under the assumption of a high $X/R_l$ ratio with line reactance $X = \omega^* L_l$ and constant voltage amplitudes.}
	\label{fig:two-source-one-load}
\end{figure}

Figure~\ref{fig:TCL_sync} depicts the frequency and angle errors following the connection of Converter~I and Converter II. 
Before the interconnection, the modulation angle of Converter I is initialized at zero with $\theta_1^*(t) = \omega^*t $, for $t<0$ and $\theta_1^*(t) = \omega^*t + \theta_1^*(0)$, for $t\geq 0$. The nominal angle of Converter II is given by $\theta^*_2(t) = \omega^*t +\theta^*_2(0),\, t\geq 0$, where $\theta^*_1(0)$ and $\theta^*_2(0)$ are the initial angle of Converter I and II at the time of interconnection $t=0$. The choice of the initial angle $\theta^*_2(0)$ can be determined as follows.
The two-converter system represented in Fig.~\ref{fig: scenarioII} can be reduced to two sources connected to one load as depicted in Fig.~\ref{fig:two-source-one-load}.

The active power at the $k$-th DC/AC converter at steady state is given by~\cite{kundur1994power} 
	\begin{align}
		P^s_{k} = \frac{V_k\,V_0}{X_{k0}} \sin{(\theta^s_k-\theta^s_0)}\,,\, k=\{1,2\}
		\label{eq: activ-pow-exp}
	\end{align}
	where $V_0\angle\theta^s_0$ is the phasor at the common node connecting the two converters and $V_k\angle\theta^s_k$ that of the switching voltage of the $k-$th DC/AC converter with $V_k=\norm{\frac{1}{2} u_k V_{dc}}$ and $X_{k0}=\omega^* L_l>~0$ is the reactance of the inductive line impedance. Thus, the active power at Converter II is given by $  P^s_{2} = \frac{V_2\,V_0}{X_{20}} \sin{(\theta^s_2-\theta^s_0)}$.
	Setting $ P^s_{2}=P^*_{2} = 0$ leads to $\theta^*_2(0)=\theta_0(0)$. From
	$ P^s_{1} = \frac{V_1\,V_0}{X_{10}} \sin{(\theta^s_1-\theta^s_0)} = P^*$, we obtain
	\begin{align}
		\theta^*_1(0) = \theta_0(0) +\arcsin(\frac{P^* X_{10}}{V_0 V_1}), 
		\label{eq: IC}
	\end{align}
	where $\arcsin(\frac{P^* X_{10}}{V_0 V_1}) = 0.006$, $\theta_1(0)$ is the initial angle of Converter I and $\theta_0(0)$ is the initial angle at node 0 depicted in Fig.~\ref{fig:two-source-one-load} obtained from a Phased-Locked-Loop (PLL) scheme at the time of interconnection.

\begin{figure}[ht!]
		\centering
		\includegraphics[width=0.85\linewidth]{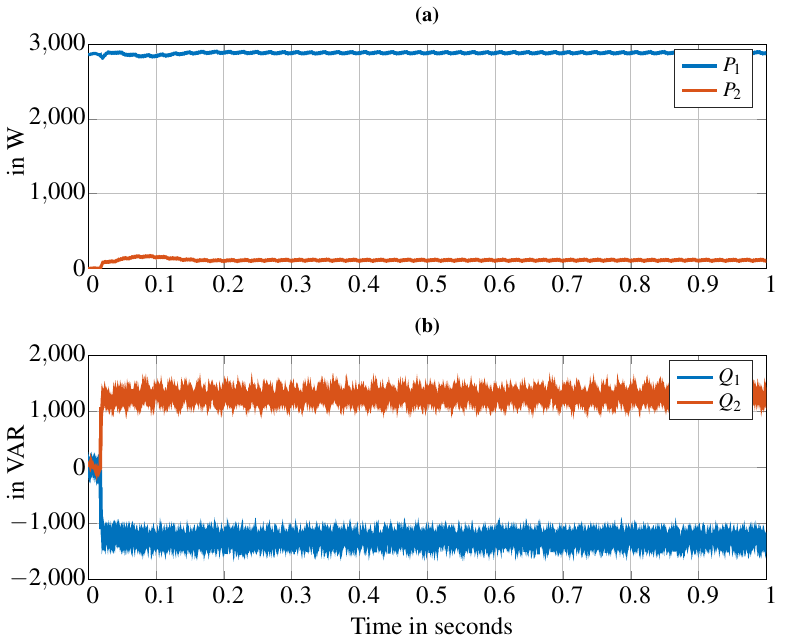}
		\caption{Active (a) and reactive (b) power $P^s_{k} = v_k^\top \,i_{o,k}$ and $Q^s_k = v_k^\top\, \mathbf{J}\, i_{o,k}$ with $P_1^*=P^*, P^*_2=0$ with $k = 1,2$ for the frequency synchronization experiment. Converter I and II in closed loop with direct angular droop with $\gamma_k=5 \cdot 10^4$ and $\alpha_k=2000$ and $k = 1,2$.}
		\label{fig:TCL_sync_power}
	\end{figure}

\begin{figure}
    	\centering
    	\includegraphics[width=\linewidth]{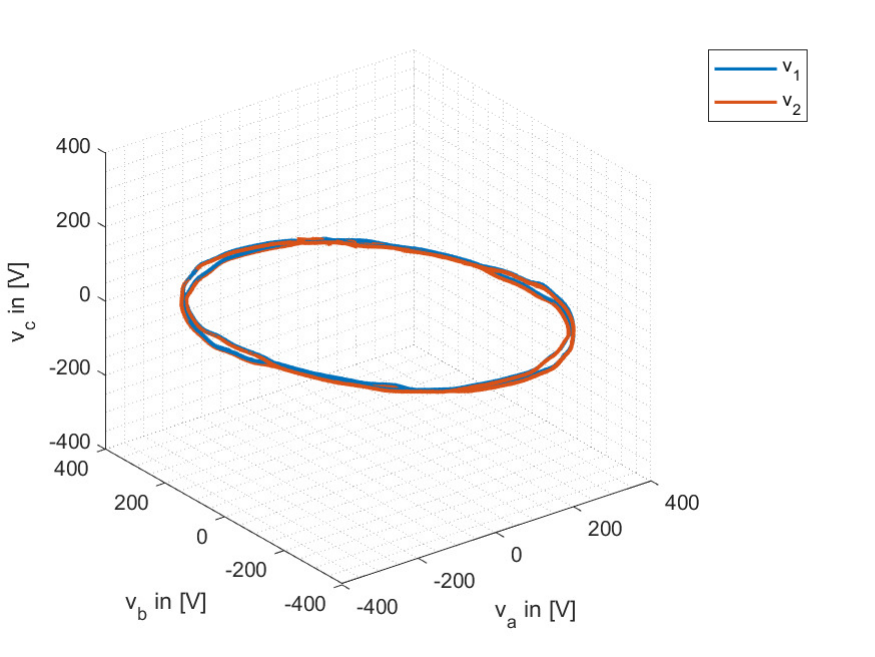}
    	\caption{Phase portrait of the periodic orbit of the three-phase output voltages $v_1$ and $v_2$ in $abc$- frame, respectively in closed-loop with             the direct angular droop control for $\gamma_k=5 \cdot 10^4$ and $\alpha_k=2000$ and $k = 1,2$ during the frequency synchronization                     experiment.}
    	\label{fig:periodic-orbit-volt}
    \end{figure}

	Our experimental results in Fig.~\ref{fig:TCL_sync} show that the modulation angle differences at steady state, i.e.,  $\theta^{s}_1-\theta^{s}_2 = 0.002$. This can be inferred as follows
	\begin{align}
		\theta_1^s-\theta_2^s &= \theta_1^*-\theta_2^* +\frac{1}{\gamma}(\delta P_1 +\delta P_2)\\
		& = \theta_1(0) - \theta_0(0) +\frac{1}{\gamma}(\delta P_1 +\delta P_2)\\
		& = \theta_0(0) +\arcsin(\frac{P^* X_{10}}{V_0 V_1}) - \theta_0(0)+\frac{1}{\gamma}(\delta P_1 +\delta P_2)\\
		& = 0.006 +\frac{1}{\gamma}(\delta P_1 +\delta P_2), 
	\end{align}
	where $\delta P_1 +\delta P_2 = -200$ and $\delta P_1 = P^*_1- P^s_1$ and $\delta P_2 = P^*_2-{ P}^s_2$.
	Thereby, the steady state angle differences remain within $[-\pi/2,\pi/2]$ (rad) and the security constraint in~\cite{jouini2022inverse} is satisfied and the two converters synchronize at nominal frequency $\omega^*$ within $0.25\,s$.
	Both converters' angles converge to their frequency synchronous steady states. 
	This corresponds to the active power of Converter I and II,  $ P^s_{1},P^s_{2}$ reaching a nearby nominal value as seen in Figure~\ref{fig:TCL_sync_power}. We note thereby that $P^*_{1}+P^*_{2} = P^*$ where $P^*$ is the total active power drawn by the load resistance. Furthermore, the converters exchange reactive power $\tilde Q^s_1<0$ and $\tilde Q^s_2>0$, where $\tilde Q^s_1+\tilde Q^s_2 = 0$ at all times $t\geq 0$. 
 In particular, in a highly inductive, Kron-reduced~\cite{6316101} power network, the reactive power is expressed at the $k-$th converter by~\cite{kundur1994power}
	\begin{align}
		\tilde Q^{s}_k = \frac{V_k}{X_{kj}}(V_k-V_j \cos(\theta_k-\theta_j)),
		\label{eq: reac-pow}
	\end{align}
where $k\neq j$ and $k=\{1,2\}$. Under the small signal approximation, i.e., $\theta_k-\theta_j\approx 0$, we obtain
	\begin{align}
		\tilde Q^{s}_k \approx \frac{V_k}{X_{kj}}(V_k-V_j).    
	\end{align}
\textcolor{rev1}{Hence, mismatches of the voltage amplitudes, $V_k\neq V_j$ (see also Fig.~\ref{fig:periodic-orbit-volt}) lead to non-zero reactive power $\tilde Q^{s}_k\neq 0$ for $k=\{1,2\}$. Such mismatches are unavoidable in hardware setups due to manufacturing tolerances of the used components and parasitic effects that occur in real-world applications.}
Finally, the phase portrait of the output voltages $v_1$ and $v_2$ in Fig.~\ref{fig:periodic-orbit-volt} represents a limit cycle of an approximate radius of $V^d$ in the phase plane which shows once again the frequency synchronization in AC voltages nearby a desired voltage amplitude $V^d$.
As a conclusion, our experiment validates local asymptotic stability result shown in~\cite{jouini2022inverse}.
%
\subsubsection{Power-sharing}
\begin{figure}
	\centering
	\includegraphics[width=0.85\linewidth]{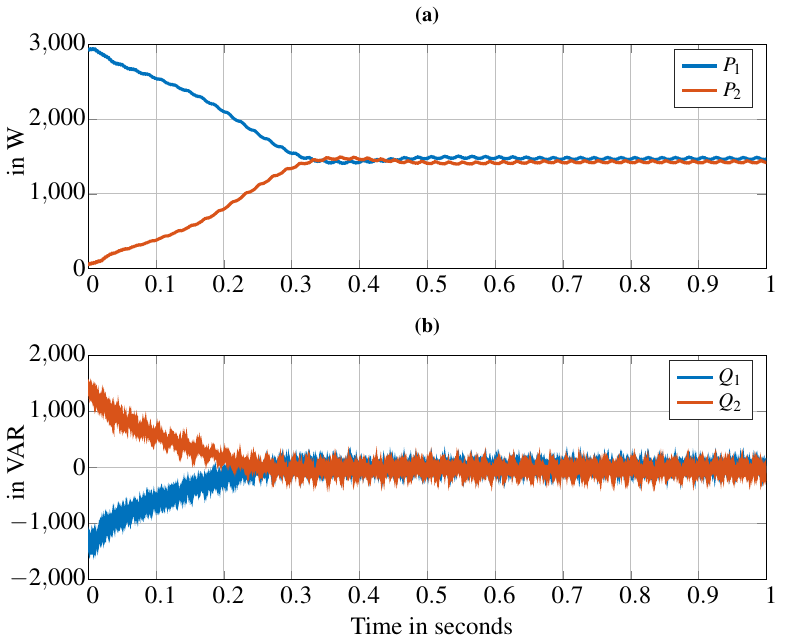}
	\caption{Active power $P_k = v_k^\top i_{o,k}$ in (a) and reactive power $Q_k = v_k^\top \mathbf{J} i_{o,k}$ in (b) following the connection of Converter II to Converter I at $t = 0$, both in  closed-loop with the direct angular droop control for $\gamma_k=500, \alpha_k=2000$ and the power ratio $r=1$ in~\eqref{eq: power-ratio} for $k = 1,2$ in the power-sharing experiment,  $P_1^*= P^*_2 = \frac{P^*}{2}$.}
	\label{fig: P-Q-pow-sharing}
\end{figure}
To achieve power-sharing among the Converters I and II, we first determine a suitable tuning for the power-to-angle droop gains $\gamma_k,\, k= 1,2$ and the nominal power ratio $P_1^*/P_2^*$, by conducting the following analysis inspired by~\cite{kolluri_2017}. To keep the analysis tractable, we assume in the remainder an inductive power network characterized by high $X/{R_l}$ ratio (see Table~\ref{tab: trans-line}), constant voltage magnitudes and neglect the output filter at each converter.
\paragraph{Choice of $(\gamma_k,P^*_k),\, k = 1,2$}
Under the small signal approximation, \eqref{eq: activ-pow-exp} can be rewritten as
\begin{align}
	\label{eq: powerflow}
	(\theta^s_k-\theta^s_0) \approx \frac{X_{k0}}{V_k\,V_0} P^s_{k}\,.
\end{align}
At steady state, the angular droop control law is given by~\eqref{eq: droop-bhv}, where
\begin{align}
	\label{eq:adroopsteadystate}
	\theta^s_k &= \theta_k^* +\frac{1}{\gamma_k} (P^*_{k}-P^s_{k}),\quad k ={1,2}.
\end{align}
Here $\theta_k^s\in\real$ and $P_{k}^s>0$ are the induced Kron-reduced steady state angle and active power at the $k-$th converter.\\
By letting $\theta^s_1-\theta^s_2 = \theta^s_1-\theta^s_0+\theta^s_0-\theta^s_2$, we obtain,
\begin{align}
	\theta^*_1-\theta^*_2 -\frac{1}{\gamma_1} (P^s_{1}-P^*_{1}) +\frac{1}{\gamma_2} (P^s_{2}-P^*_{2}) = \frac{X_{10}}{V_1\,V_0} P^s_{1} -\frac{X_{20}}{V_2\,V_0} P^s_{2}.
	\label{eq: angle-diff}
\end{align}
By reordering the terms in~\eqref{eq: angle-diff}, we arrive at
\begin{align}
	\theta^*_1&-\theta^*_2 +\frac{1}{\gamma_1} P_{1}^* - \frac{1}{\gamma_2} P_{2}^*=\bigg(\frac{1}{\gamma_1}+\frac{X_{10}}{V_1\,V_0}\bigg) P^s_{1} -\bigg(\frac{1}{\gamma_2}+\frac{X_{20}}{V_2\,V_0}\bigg) P^s_{2}.
\end{align}
From~\eqref{eq: IC}, we have that $\theta^*_1(t)-\theta^*_2(t)=\theta_1(0)-\theta_0(0)= 0.006$. Therefore, if we select the power-to-angle-droop gain $\gamma_k>0$ such that,
\begin{align}
	\label{eq: gammangular cdt}
	\gamma_k \ll \frac{V_k V_0}{X_{k0}},\quad k=1,2
\end{align}
holds, it yields that $\frac{1}{\gamma_1} P_{1}^* - \frac{1}{\gamma_2} P_{2}^* \approx\frac{1}{\gamma_1} P^s_{1} -\frac{1}{\gamma_2} P^s_{2}$, and we deduce that for an active power ratio defined by, 
\begin{align}
	\label{eq: power-ratio}
	r := \frac{P^*_{1}}{P^*_{2}} = \frac{\gamma_{1}}{\gamma_{2}}.    
\end{align} Therefore, $\frac{P^s_{1}}{P^s_{2}} \approx r$ and the power-sharing between the two converters is guaranteed. Finally, for our experimental setup, the condition~\eqref{eq: gammangular cdt} can be rewritten as
\begin{align}
	\label{eq: gammangular cdt-exp}
	\gamma_k \ll 4.8\cdot 10^5, 
\end{align}
with $V_1 = V_2 = V^d, V_0\approx V^d $ and $X_{10} = X_{20} = \omega^* L_l$. To achieve~\eqref{eq: gammangular cdt-exp}, we select $\gamma_1 =\gamma_2 =  500$ with $r=1$ throughout the power-sharing experiment.
\paragraph{Discussion}
Figure~\ref{fig: P-Q-pow-sharing} shows the experimental results of the power-sharing experiment. The security constraint in~\cite{jouini2022inverse} is here again satisfied with $\norm{\theta^s_1-\theta^s_2}<0.1$.
For $\gamma_1 = \gamma_2 = 500$, power-sharing is guaranteed at steady state, where $P^s_{k} \approx P^*/2$ for $k = 1,2$. This corresponds to zero reactive power at steady state. As expected, the two converters synchronize in frequency and their angles converge respectively to frequency synchronous steady states within $0.5\,$s. Thus, our experiment validates the power-sharing capabilities of the angular-droop controlled DC/AC converter system, one of the most important plug and play properties for converter control design  in power networks.   
\section{Conclusion}
\label{sec:concl} 
We demonstrate the grid-forming properties of the angular droop control in two scenarios embedded in a hardware experiment setup. For this we provide traceable analysis and solutions to the challenges that arise from virtual angle control in real-world applications. First, the discretized control law is rewritten to conform with the restrictions posed by hardware implementation. Second, the susceptibility to clock drifts in multi-converter settings is analyzed and resolved by distributing a common master clock. Using a single converter setup, the black start capabilities of angular droop control as well as the capability to withstand sudden load changes while returning to zero frequency deviation is shown. With the extension to a multi-converter setup the frequency synchronization as well as load sharing properties are demonstrated and tuning guidelines are established.
Our future work aims to extend the angular droop control with voltage regulation, therefore relaxing the assumption on constant AC voltage amplitude. Furthermore, the interoperability of angular droop with other control strategies needs to be considered. \textcolor{rev1}{For prospective industrial-microgrid applications, the influence of nonlinear loads as well as the extension to larger multi-converter setup will be investigated.}
\subsection*{Author Contributions}
 \textbf{Taouba Jouini:} conceptualization, data curation, formal analysis, investigation, methodology, visualization, writing - original draft, writing - review \& editing
 \textbf{Jan Wachter:} conceptualization, data curation, formal analysis, investigation, methodology, visualization, writing - original draft, writing - review \& editing
 \textbf{Sophie An:} conceptualization, formal analysis, investigation, methodology, writing - original draft
 \textbf{Veit Hagenmeyer:} conceptualization, funding acquisition, supervision, writing - review \& editing
 
 \subsection*{Conflict of Interest}
 The authors declare no conflict of interest.
  
 \subsection*{Data Availability Statement}
The data that support the findings of this study are available from the corresponding author upon reasonable request.
\bibliographystyle{IEEEtranTIE}
\bibliography{root.bib}\ 

@article{SIMPSONPORCO20132603,
	title = {Synchronization and power sharing for droop-controlled inverters in islanded microgrids},
	journal = {Automatica},
	volume = {49},
	number = {9},
	pages = {2603--2611},
	year = {2013},
	issn = {0005-1098},
	doi = {https://doi.org/10.1016/j.automatica.2013.05.018},
    author = {John W. Simpson-Porco and Florian D{\"o}rfler and Francesco Bullo}
}

@article{wiegel2022smart,
  title={Smart Energy System Control Laboratory -- a fully-automated and user-oriented research infrastructure for controlling and operating smart energy systems},
  author={Wiegel, Friedrich and Wachter, Jan and Kyesswa, Michael and Mikut, Ralf and Waczowicz, Simon and Hagenmeyer, Veit},
  journal={at-Automatisierungstechnik},
  volume={70},
  number={12},
  pages={1116--1133},
  year={2022},
  publisher={De Gruyter (O)}
}

@ARTICLE{8638531,
	
	author={Colombino, Marcello and Gro\ss, Dominic and Brouillon, Jean-Sébastien and D{\"o}rfler, Florian},
	
	journal={IEEE Transactions on Automatic Control}, 
	
	title={Global Phase and Magnitude Synchronization of Coupled Oscillators With Application to the Control of Grid-Forming Power Inverters}, 
	
	year={2019},
	
	volume={64},
	
	number={11},
	
	pages={4496-4511},
	
	doi={10.1109/TAC.2019.2898549}}

@inproceedings{seo2019dispatchable,
  title={Dispatchable virtual oscillator control for decentralized inverter-dominated power systems: Analysis and experiments},
  author={Seo, Gab-Su and Colombino, Marcello and Subotic, Irina and Johnson, Brian and Gro{\ss}, Dominic and D{\"o}rfler, Florian},
  booktitle={2019 IEEE Applied Power Electronics Conference and Exposition (APEC)},
  pages={561--566},
  year={2019},
  organization={IEEE}
}

@inproceedings{jouini2022inverse,
  title={Inverse optimal control for angle stabilization in converter-based generation},
  author={Jouini, Taouba and Rantzer, Anders and Tegling, Emma},
  booktitle={2022 American Control Conference (ACC)},
  pages={4945--4950},
  year={2022}
}

@book{kundur1994power,
  title={Power system stability and control},
  author={Kundur, Prabha and Balu, Neal J and Lauby, Mark G},
  volume={7},
  year={1994},
  publisher={McGraw-Hill New York}
}

@ARTICLE{9616402,
	author={Jouini, Taouba and Sun, Zhiyong},
	journal={IEEE Transactions on Control of Network Systems}, 
	title={Frequency Synchronization of a High-Order Multiconverter System}, 
	year={2022},
	volume={9},
	number={2},
	pages={1006--1016},
	doi={10.1109/TCNS.2021.3128493}}

@ARTICLE{kolluri_2017,
  author={Kolluri, Ramachandra Rao and Mareels, Iven and Alpcan, Tansu and Brazil, Marcus and de Hoog, Julian and Thomas, Doreen Anne},
  journal={IEEE Transactions on Power Systems}, 
  title={Power Sharing in Angle Droop Controlled Microgrids}, 
  year={2017},
  volume={32},
  number={6},
  pages={4743--4751},
  doi={10.1109/TPWRS.2017.2672569}}

@ARTICLE{7781662,
  author={Schiffer, Johannes and Hans, Christian A. and Kral, Thomas and Ortega, Romeo and Raisch, Jörg},
  journal={IEEE Transactions on Industrial Electronics}, 
  title={Modeling, Analysis, and Experimental Validation of Clock Drift Effects in Low-Inertia Power Systems}, 
  year={2017},
  volume={64},
  number={7},
  pages={5942--5951},
  doi={10.1109/TIE.2016.2638805}}

@ARTICLE{clocksynch,
  journal={IEC 61588:2009(E)},
  author = {IEC},
  title={{IEC/IEEE} International Standard - Precision Clock Synchronization Protocol for Networked Measurement and Control Systems}, 
  year={2009},
  pages={1--292},
  doi={10.1109/IEEESTD.2009.4839002}}

@manual{fn:PEB8038,
  title = {Half-bridge SiC power module},
  organization  = "Imperix",
  note = {Rev.C / March 2021},
  url = {https://imperix.com/wp-content/uploads/document/PEB8038.pdf}
}

@manual{B-Box,
	organization  = "Imperix",
	title         = "B-Box RCP rapid prototyping controller",
	note          = {Rev. 11/08/21},
    url           = {https://imperix.com/wp-content/uploads/document/B-Box_Datasheet.pdf}
}

@manual{passiv-rack,
  title = {Passive filters box},
  organization  = "Imperix",
  note          = {Rev. D / March, 2021},
  url = {https://imperix.com/wp-content/uploads/document/Passives_Rack.pdf}
}

@manual{imperixsynch,
	organization  = "Imperix",
	title         = "Real Sync",
	url            = {https://imperix.com/technology/distributed-modulation/}
}

@misc{freq_reg,
    author= {{Commission of European Union (EU)}},
    title = {{Commission Regulation (EU) 2017/1485 of 2 August 2017 establishing a guideline on electricity transmission system operation}},
    url = {https://eur-lex.europa.eu/legal-content/EN/TXT/?uri=CELEX:02017R1485-20210315}
}

@ARTICLE{6316101,
  author={D{\"o}rfler, Florian and Bullo, Francesco},
  journal={IEEE Transactions on Circuits and Systems I: Regular Papers}, 
  title={Kron Reduction of Graphs With Applications to Electrical Networks}, 
  year={2013},
  volume={60},
  number={1},
  pages={150--163},
  doi={10.1109/TCSI.2012.2215780}}

@article{ARGHIR2018273,
title = {Grid-forming control for power converters based on matching of synchronous machines},
journal = {Automatica},
author = {Catalin Arghir and Taouba Jouini and Florian D{\"o}rfler},
volume = {95},
pages = {273--282},
year = {2018},
issn = {0005-1098},
doi = {https://doi.org/10.1016/j.automatica.2018.05.037}
}

@article{JOUINI2016192,
title = {Grid-Friendly Matching of Synchronous Machines by Tapping into the DC Storage},
journal = {IFAC-PapersOnLine},
author = {Taouba Jouini and Catalin Arghir and Florian D{\"o}rfler},
volume = {49},
number = {22},
pages = {192--197},
year = {2016},
issn = {2405-8963},
doi = {https://doi.org/10.1016/j.ifacol.2016.10.395}
}

@article{johnson2015synthesizing,
  title={Synthesizing virtual oscillators to control islanded inverters},
  author={Johnson, Brian B and Sinha, Mohit and Ainsworth, Nathan G and D{\"o}rfler, Florian and Dhople, Sairaj V},
  journal={IEEE Transactions on Power Electronics},
  volume={31},
  number={8},
  pages={6002--6015},
  year={2015},
  publisher={IEEE}
}

@ARTICLE{9429728,
  author={Jouini, Taouba and Rantzer, Anders},
  journal={IEEE Control Systems Letters}, 
  title={On Cost Design in Applications of Optimal Control}, 
  year={2022},
  volume={6},
  number={},
  pages={452-457},
  doi={10.1109/LCSYS.2021.3079642}}

@INPROCEEDINGS{7171084,

  author={Sinha, Mohit and D{\"o}rfler, Florian and Johnson, Brian B. and Dhople, Sairaj V.},

  booktitle={2015 American Control Conference (ACC)}, 

  title={Virtual Oscillator Control subsumes droop control}, 

  year={2015},

  volume={},

  number={},

  pages={2353--2358},

  doi={10.1109/ACC.2015.7171084}}

@ARTICLE{6692879,
  author={Johnson, Brian B. and Dhople, Sairaj V. and Hamadeh, Abdullah O. and Krein, Philip T.},
  journal={IEEE Transactions on Power Electronics}, 
  title={Synchronization of Parallel Single-Phase Inverters With Virtual Oscillator Control}, 
  year={2014},
  volume={29},
  number={11},
  pages={6124--6138},
  doi={10.1109/TPEL.2013.2296292}}

@INPROCEEDINGS{5275987,
  author={Majumder, Ritwik and Ghosh, Arindam and Ledwich, Gerard and Zare, Firuz},
  booktitle={2009 IEEE Power \& Energy Society General Meeting}, 
  title={Angle droop versus frequency droop in a voltage source converter based autonomous microgrid}, 
  year={2009},
  volume={},
  number={},
  pages={1-8},
 }

@article{jouini2023a,
title = {Input and state constrained inverse optimal control with application to power networks},
author = {Taouba Jouini and Zhiyong Sun and Venkatraman Renganathan and Veit Hagenmeyer},
journal = {IFAC-PapersOnLine},
volume = {56},
number = {2},
pages = {5451-5456},
year = {2023},
note = {22nd IFAC World Congress},
issn = {2405-8963},
doi = {https://doi.org/10.1016/j.ifacol.2023.10.196}
}

@INPROCEEDINGS{jouini2023b,
  author={Jouini, Taouba and Sun, Zhiyong and Hagenmeyer, Veit},
  booktitle={2023 IEEE Conference on Control Technology and Applications (CCTA)}, 
  title={Tuning of discrete-time angular droop controllers}, 
  year={2023},
  volume={},
  number={},
  pages={741-745},
  doi={10.1109/CCTA54093.2023.10252209}}

@phdthesis{tj_phd,
title = "Network Synchronization and Control Based on Inverse Optimality: A Study of Inverter-Based Power Generation",
author = "Taouba Jouini",
year = "2021",
month = dec,
day = "15",
language = "English",
isbn = "978-91-8039-090-3",
publisher = "Department of Automatic Control, Lund University",
number = "1134",
type = "Doctoral Thesis",
school = "Department of Automatic Control",

}

@article{wachter2024,
  title = {Survey of {{Real-World Grid Incidents}}--{{Opportunities}}, {{Arising Challenges}} and {{Lessons Learned}} for the {{Future Converter Dominated Power System}}},
  author = {Wachter, Jan and Gr{\"o}ll, Lutz and Hagenmeyer, Veit},
  year = {2024},
  journal = {IEEE Open Journal of Power Electronics},
  volume = {5},
  pages = {50--69},
  issn = {2644-1314},
  doi = {10.1109/OJPEL.2023.3343167}
}

@article{sajadi2023,
  title = {Dynamics and {{Stability}} of {{Power Systems With High Shares}} of {{Grid-Following Inverter-Based Resources}}: {{A Tutorial}}},
  shorttitle = {Dynamics and {{Stability}} of {{Power Systems With High Shares}} of {{Grid-Following Inverter-Based Resources}}},
  author = {Sajadi, Amirhossein and Ra{\~n}ola, Jo Ann and Kenyon, Rick Wallace and Hodge, Bri-Mathias and Mather, Barry},
  year = {2023},
  journal = {IEEE Access},
  volume = {11},
  pages = {29591--29613},
  issn = {2169-3536},
  doi = {10.1109/ACCESS.2023.3260778}
}

@article{dorfler2023a,
  title = {Control of {{Low-Inertia Power Systems}}},
  author = {D{\"o}rfler, Florian and Gro{\ss}, Dominic},
  year = {2023},
  month = may,
  journal = {Annual Review of Control, Robotics, and Autonomous Systems},
  volume = {6},
  number = {1},
  pages = {415--445},
  issn = {2573-5144, 2573-5144},
  doi = {10.1146/annurev-control-052622-032657}
}

@article{chakraborty2024,
  title = {Seamless {{Transition}} of {{Critical Infrastructures Using Droop-Controlled Grid-Forming Inverters}}},
  author = {Chakraborty, Soham and Patel, Sourav and Saraswat, Govind and Maqsood, Atif and Salapaka, Murti V.},
  year = {2024},
  month = feb,
  journal = {IEEE Transactions on Industrial Electronics},
  volume = {71},
  number = {2},
  pages = {1535--1546},
  issn = {0278-0046, 1557-9948},
  doi = {10.1109/TIE.2023.3253946}
}

@manual{currentsensor,
	organization  = "Imperix",
	title         = "50A DIN rail-mountable current sensor",
    note          = {Datasheet},
    url           = {https://imperix.com/wp-content/uploads/document/DIN-50A.pdf}
}

@manual{voltagesensor,
	organization  = "Imperix",
	title         = "800V DIN rail-mountable voltage sensor",
    note          = {Datasheet},
    url           = {https://imperix.com/wp-content/uploads/document/DIN-800V.pdf}
}

@INPROCEEDINGS{Golshani2021,
  author={Golshani, M. and Wilson, D. and Norris, S. and Cowan, I. and Rahman, Md H. and Marshall, B.},
  booktitle={The 17th International Conference on AC and DC Power Transmission (ACDC 2021)}, 
  title={Application of phasor-based functionality to {HVDC} control in reduced system strength}, 
  year={2021},
  volume={2021},
  number={},
  pages={44-49},
  doi={10.1049/icp.2021.2442}}

@article{Fank2024,
author = {Fank, Daniel and Renner, Herwig},
title = {Deployment of a full-size converter utilised hydropower plant to enhance inter-area oscillation damping},
journal = {IET Generation, Transmission \& Distribution},
volume = {18},
number = {10},
pages = {1992-2005},
doi = {https://doi.org/10.1049/gtd2.13165},
year = {2024}
}

@article{wang2020iet,
author = {Wang, Tong and Yang, Jing and Padhee, Malhar and Bi, Jingtian and Pal, Anamitra and Wang, Zengping},
title = {Robust, coordinated control of SSO in wind-integrated power system},
journal = {IET Renewable Power Generation},
volume = {14},
number = {6},
pages = {1031-1043},
doi = {https://doi.org/10.1049/iet-rpg.2019.0410},
year = {2020}
}

@INPROCEEDINGS{Strunk2024,
  author={Strunk, Robin and Sourkounis, Pieris and Mertens, Axel},
  booktitle={2024 IEEE 15th International Symposium on Power Electronics for Distributed Generation Systems (PEDG)}, 
  title={Impedance-Based Stability Analysis of Grid-Forming Inverters with Virtual Impedance or Angle Droop for Improved Robustness}, 
  year={2024},
  volume={},
  number={},
  pages={1-6},
  doi={10.1109/PEDG61800.2024.10667448}}

@article{Xu2021,
author = {Xu, Tao and Zhou, Jiaxin and Liang, Lemeng and Wu, Yuhan and Cai, Shuqi and Liu, Zuozheng and Li, Peng and Yu, Li},
title = {Consensus active power sharing for islanded microgrids based on distributed angle droop control},
journal = {IET Renewable Power Generation},
volume = {15},
number = {13},
pages = {2826-2839},
doi = {https://doi.org/10.1049/rpg2.12210},
year = {2021}
}
\appendix
\label{ap:appendixA}
\section{High-order converter modeling  and control}
\label{sec:high-order-model}
This section details how the angular droop control law \eqref{eq: angular droop} is implemented on the experiment hardware.
With some abuse of notation, we omit throughout subsequent sections the subscript $k$ to denote a quantity $z_k$ of the $k-$th converter.
\subsection{Modeling and control DC/AC converter}
We start by relaxing our modeling assumptions from Section~\ref{sec:Understand-adroop-ctrl} towards a hardware experiment in two directions:
\begin{itemize}
    \item First, we include a DC power supply and a DC/DC converter relying on vector control consisting of cascaded voltage and current control loops put in series behind the DC/AC converter to provide adequate DC-link voltage.
    \item Second, even though the derivation of the angular droop control ignores the internal dynamics of the converter, we present a sufficiently detailed, high-order model of the DC/AC converter. Based on it, we suggest a direct and indirect method to implement the angular droop control~\eqref{eq: angular droop} by modulation control.
\end{itemize}
Hereby, we further assume that all AC voltage amplitudes are constant and at nominal.
\subsubsection{Modeling DC/AC converter}
We consider a three-phase, averaged and balanced DC/AC converter as shown in Fig.~\ref{fig: DCAC_converter} and described in $abc-$frame~\cite{kundur1994power, 9616402},
\begin{equation}
	\begin{aligned}
		\label{eq: open-loop}
		L \,\dot i&= -R \,i + \frac{1}{2}  \overline u \, \,V_{dc} -v , \\
		C \,\dot v&= -G\, v +i- i_{o},   
	\end{aligned}
\end{equation}
where $V_{dc}$ denotes the DC side capacitor voltage. On the AC side, let $i\in\real^3$ denote the inductance current
and $v\in\real^3$ the output voltage. The filter resistance and
inductance are specified by $R > 0$ and $L > 0$, respectively. The capacitor $C > 0$ is set in parallel with the load conductance
$G > 0$ to ground. The DC/AC converter is connected to the AC network, where $i_{o}\in\real$ is the output current flowing into the network. 
Note that the pulse width modulation signal $\overline u \in[-1 , 1]$ relates to the converter duty cycle $d_{AC}\in[0,1]$ via
\begin{align}
	\label{AC-duty-cycle}
	d_{AC} = \frac{1}{2}+\frac{\overline u}{2},
\end{align}
where $\overline u$ represents the main input to the DC/AC converter.
%
\subsubsection{Direct implementation}
\label{subsec: control}
First, we define the active power $P=v^\top i_{o}$, and the nominal steady state $ P^*=v^{*\top} i_{o}^*$.
For the direct implementation of the angular droop control we propose
\begin{subequations}
	\label{eq: angular droop-impl}
	\begin{align}
		\dot{\theta} &=-\frac{1}{2\alpha}\left(\gamma (\theta-\theta^*)+( P- P^*) \right)+\omega^*, \label{eq: angl-droop-impl} \\ 
		\overline u&=A \begin{bmatrix}
			\sin(\theta) \\ \sin(\theta-\frac{2\pi}{3})   \\ \sin(\theta+\frac{2\pi}{3}) 
		\end{bmatrix}, 
	\end{align}
\end{subequations}
where $0<A<1$ is the amplitude of the modulation signal $\overline u$. Fig.~\ref{fig: direct_ctrl} depicts a summarizing block diagram of the direct implementation of the angular droop control~\eqref{eq: angular droop}. Note that in~\eqref{eq: angular droop-impl}, the angular droop control increments the converter's state with a virtual angle $\theta$ through the modulation signal~$\bar{u}$.
\begin{figure}[ht!]
	\centering
	\begin{circuitikz}[scale=0.5, smallR/.style={european resistor, resistors/scale=0.5},node distance=2.0cm,font=\footnotesize,thick]
		\draw node (start) at (0,0) {};
		\node[draw, circle, right = 0.5cm of start, minimum size = 0.3cm, inner sep = 0pt] (sumP) {};
		\node[draw, circle, right = 0.5cm of sumP, minimum size = 0.3cm, inner sep = 0pt] (sumG) {};
		\node (GainGamma) [textbox, below = 0.5cm of sumG, minimum size = 0.3cm] {$\gamma$};
		\node[draw, circle, right = 0.5cm of GainGamma, minimum size = 0.3cm, inner sep = 0pt] (sumT) {};
		\node (GainAlpha) [textbox, right = 0.5cm of sumG, minimum size = 0.3cm] {$-\frac{1}{2 \alpha}$};
		\node[draw, circle, right = 0.5cm of GainAlpha, minimum size = 0.3cm, inner sep = 0pt] (sum) {};
		\node (int) [textbox, right = 0.5cm of sum,minimum size = 0.3cm] {$\frac{1}{s}$};
		\node (intDOT) [right = 0.2cm of int] {};
		\filldraw (intDOT) circle(4pt);
		\node (PWM) [textbox, right = 0.8cm of int] {Mod};
		\node (inv) [right = 0.5cm of PWM] {};
		\draw[->,forestgreen] (start) -- node[pos=0.5, above] {$P$} (sumP);
		\draw[<-,darkblue] (sumP) -- node[pos=0.1, right] {\color{black}{$-$}} ++(0,-1) -- node[pos=0.1, left] {$P^*$} ++(0,-0.5);
		\draw[->] (sumP) -- node[pos=0.5, above] {} (sumG);
		\draw[->] (sumG) -- node[pos=0.5, above] {} (GainAlpha);
		\draw[->] (GainAlpha) -- node[pos=0.5, above] {} (sum);
		\draw[<-,darkblue] (sum) -- node[pos=0.1, right] {} ++(0,1) -- node[pos=0.5, left] {$\omega^*$} ++(0,0.5);
		\draw[->] (sum) -- node[pos=0.5, above] {$\dot\theta$} (int);
		\draw[->] (int) -- node[pos=0.3, above] {$\theta$} (PWM);
		\draw[->] (PWM) -- node[pos=0.5, above] {$\bar u$} (inv);
		\draw[->] (intDOT.center) |- node[pos=0.5, above] {} (sumT);
		\draw[->] (sumT) -- node[pos=0.5, above] {} (GainGamma);
		\draw[->] (GainGamma) -- node[pos=0.5, above] {} (sumG);
		\draw[<-,darkblue] (sumT) -- node[pos=0.1, left] {\color{black}{$-$}} ++(0,-1) -- node[pos=0.1, right] {$\theta^*$} ++(0,-0.5);
	\end{circuitikz}
	\caption{Implementation of the angular droop for the DC/AC converter via direct control of the modulation signal $\bar u$.}
	\label{fig: direct_ctrl}
\end{figure}
%
%
\subsubsection{Indirect Implementation}
\label{para: indirect-impl}
We propose an indirect implementation of the angular droop control that relies on cascaded control. In particular, the indirect implementation entails inner voltage and current control loop according to a cascaded architecture. After a Park transformation $\mathcal{P}(\theta_{dq})$ with angle $\theta_{dq}(t):=\theta(t)$ and given the reference voltage $v^d$ in $abc$-frame~\cite{kundur1994power},  
\begin{align}
	\label{eq: vol_ref}
	v^{d}(\theta) = V^{*} \begin{bmatrix}
		\sin(\theta) \\ \sin(\theta-\frac{2\pi}{3}) \\ \sin(\theta+\frac{2\pi}{3})
	\end{bmatrix},
\end{align}
with $V^{*}>0$ the reference voltage amplitude and $\theta$ the angle given by~\eqref{eq: cl-dyn}, the tracking of the reference voltage~\eqref{eq: vol_ref} is achieved via  cascaded voltage and current loops implementing proportional-integral (PI) controllers. Thereby, the outer voltage loop generates a reference current signal
\begin{align}
	i_{dq}^d = Y \, v_{dq} + i_{o,dq} -k_{VP} (v_{dq}-v_{dq}^d)- k_{VI} \int_ 0^t \big(v_{dq}(\tau)-v_{dq}^d \big) \dd \tau,
	\label{eq: ref-current}
\end{align}
where $k_{VP}, k_{VI}~>0$ are the control gains and $Y= G + C \mathbf{J} \omega^*$. To track the reference current~\eqref{eq: ref-current}, we design an inner current loop based on PI control using the switching voltage $v^d_m = \frac{1}{2} \overline u_{dq} \, V_{dc} $ as follows
\begin{align}
	v_m^d =  Z \, i_{dq}+ v_{dq} -k_{IP} (i_{dq}-i^d_{dq}) - k_{II} \int_ 0^t  (i_{dq}(\tau)-i_{dq}^d) \, \dd \tau, 
\end{align}
where $k_{IP}, k_{II} >0$ are control gains and $Z = R + L\, \mathbf{J}\, \omega^*$. By applying the inverse Park transformation $\mathcal{P}^{-1}(\theta_{dq})$, we recover the modulation input $\overline u$ in $abc-$frame as follows, 
\begin{align}
	\overline u(\theta) = 2\; \frac{\mathcal{P}^{-1}(\theta_{dq})(v^d_m)}{V_{dc}}.
	\label{eq: indirect-ctrl}
\end{align}
It is noteworthy that, the pairs $(k_{VP}, k_{VI})$ and $(k_{IP}, k_{II})$ are chosen to guarantee time-scale separation, where the current control loop is faster than the voltage control loop. Fig.~\ref{fig: indirect_ctrl} summarizes the overall scheme of the indirect implementation based on cascaded control.
\begin{figure}[h!]
	\centering
	\begin{circuitikz}[scale=0.5, smallR/.style={european resistor, resistors/scale=0.5},node distance=2.0cm,font=\footnotesize,thick]
		\ctikzset{
		}
		\draw node (start) at (0,0) {};
		\node (Adroop) [textbox, right = 0.3cm of start, minimum height = 1.5cm] {A-droop};
		\node (Vctrl) [textbox, right = 0.4cm of Adroop, minimum height = 1.0cm] {PI-VC};
		\node (Ictrl) [textbox, right = 0.4cm of Vctrl, minimum height = 1.0cm] {PI-CC};
		\node (Trafo) [textbox, right = 0.4cm of Ictrl] {$\mathcal{P}^{-1}_{dq}$};
		\node (PWM) [textbox, right = 0.3cm of Trafo] {PWM};
		\node (inv) [ right = 0.4cm of PWM] {};
		\draw[<-,forestgreen] (Adroop.140) -- node[pos=0.5, above] {$P$} ++(-1,0);
		\draw[<-,darkblue] (Adroop.180) -- node[pos=0.5, above] {$P^*$} ++(-1,0);
		\draw[<-,darkblue] (Adroop.220) -- node[pos=0.5, above] {$\theta^*$} ++(-1,0);
		\draw[<-,darkblue] (Adroop.220) -- node[pos=0.5, above] {$\theta^*$} ++(-1,0);
		\draw[->] (Adroop) -- node[pos=0.5, above] {$\theta$} (Vctrl);
		\draw[<-,darkblue] (Vctrl.90) -- node[pos=0.5, left] {$V^*$} ++(0,1);
		\draw[<-,forestgreen] (Vctrl.270) -- node[pos=0.5, left] {$v_{dq}$} ++(0,-1);
		\draw[->] (Vctrl) -- node[pos=0.5, above] {$i_{dq}^d$} (Ictrl);
		\draw[<-,forestgreen] (Ictrl.270) -- node[pos=0.5, left] {$i_{dq}$} ++(0,-1);
		\draw[->] (Vctrl) -- node[pos=0.5, above] {$i_{dq}^d$} (Ictrl);
		\draw[->] (Ictrl) -- node[pos=0.5, above] {$v_{m}^d$} (Trafo);
		\draw[->] (Trafo) -- node[pos=0.5, above] {} (PWM);
		\draw[->] (PWM) -- node[pos=0.5, above] {$\bar u$} (inv);
	\end{circuitikz}
	\caption{Indirect cascaded control of the angular droop after $dq$-transformation of the DC/AC converter. PI-VC and PI-CC denote the proportional integral (PI) voltage and current controller, respectively.}
	\label{fig: indirect_ctrl}
\end{figure}
\color{rev2}
\subsection{Modeling and control DC/DC boost converter}
\subsubsection{Modeling DC/DC boost converter}
\begin{figure}[h!]
\centering
	\includegraphics[width=0.8\linewidth]{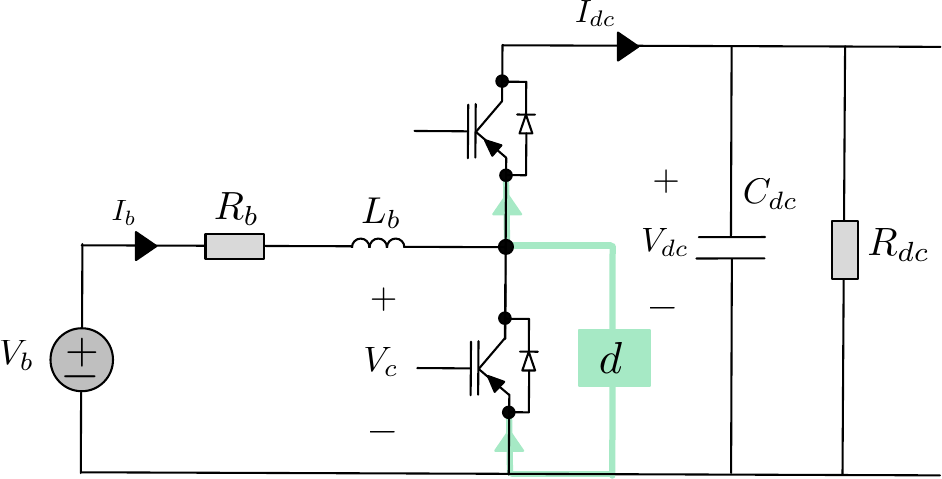}
	\caption{A schematic representation of the boost converter consisting of a half-bridge module and an inductance $L_b$ with the parasitic resistance $R_b$.}
	\label{fig: boost}
\end{figure}
Fig.~\ref{fig: boost} depicts the boost converter in our experimental setup. Observe that the two-switch configuration is due to the available half-bridge modules. However, only the lower switch is actuated to perform the desired control action. The upper switch is actuated complementary to reduce the losses through the parallel diode. 
The boost converter is modeled by following ordinary differential equations,
\begin{equation}
\begin{aligned}
L_b \dot I_b &= -R_b\, I_b + V_b -V_c \,,\\
C_{dc} \dot V_{dc} &= - G_{dc} V_{dc} +I_{dc},\quad  I_{dc}=\frac{V_b I_b}{V_{dc}}\,.
\end{aligned}
\label{eq: boost-eq}
\end{equation}

In~\eqref{eq: boost-eq}, we denote by $I_b\in\real$ the current flowing out of the DC supply and by $V_b$ the DC supply voltage. The conductance $G_{dc}=1/R_{dc}>0$ models the parasitic losses on the DC side. The inductance is represented by $L_b$ and the DC capacitance is given by $C_{dc}>0$.
Additionally, $V_{dc}\in\real$ represents the voltage across the DC capacitor and $V_c\in\real$ is the voltage controlled directly by the duty cycle $d\in[0,1]$  via the relationship
\begin{align}
\label{eq: ctrl-input}
V_c = (1-d)\cdot V_{dc}\in\real.
\end{align}
Note that the duty cycle $d$ in~\eqref{eq: ctrl-input} represents the main control input to the boost converter.
\subsubsection{Boost converter}
\label{subsubsec: control_boost}
\begin{figure}[h!]
\centering
	\begin{circuitikz}[scale=0.5, smallR/.style={european resistor, resistors/scale=0.5},node distance=2.0cm,font=\footnotesize,thick]
		\ctikzset{
		}
		\draw node (start) at (0,0) {};
		\node[draw, circle, right = 0.5cm of start, minimum size = 0.3cm, inner sep = 0pt] (sum1) {};
		\node (Vctrl) [textbox, right = 0.4cm of sum1, minimum height = 1.0cm] {PI-VC};
        \node[draw, circle, right = 0.6cm of Vctrl, minimum size = 0.3cm, inner sep = 0pt] (sum2) {};
		\node (Ictrl) [textbox, right = 0.4cm of sum2, minimum height = 1.0cm] {PI-CC};
		\node (PWM) [textbox, right = 0.6cm of Ictrl] {PWM};
		\node (inv) [ right = 0.4cm of PWM] {};
        \draw[->,darkblue] (start) -- node[pos=0.5, above] {$V^*_{dc}$} (sum1);
        \draw[<-,forestgreen] (sum1) -- node[pos=0.5, right] {$-$} ++(0,-0.5) -- node[pos=0.5, left] {$V_{dc}$} ++(0,-1);
        \draw[->] (sum1) -- node[pos=0.5, left] {} (Vctrl);
        \draw[->] (Vctrl) -- node[pos=0.5, above] {$I_b^*$} (sum2);
        \draw[<-,forestgreen] (sum2) -- node[pos=0.5, right] {$-$} ++(0,-0.5) -- node[pos=0.5, left] {$I_{b}$} ++(0,-1);
        \draw[->] (sum2) -- node[pos=0.5, left] {} (Ictrl);
        \draw[->] (Ictrl) -- node[pos=0.5, above] {$d$} (PWM);
        \draw[->] (PWM) -- ++(2,0);
        \end{circuitikz}
	\caption{Summary of the vector control of the boost converter using as main input the duty cycle $d$ in~\eqref{eq: DC-duty-cycle}.}
	\label{fig: boost-ctrl}
\end{figure}
The cascaded control architecture to regulate the DC-bus voltage of the boost converter exploits the differential equations~\eqref{eq: boost-eq} and is summarized in Fig.~\ref{fig: boost-ctrl}. In particular, an outer loop regulates the DC capacitor voltage $V_{dc}$ at a nominal value $V_{dc}^*>V_b>0$ by specifying a reference current $I^d_b\in\real$ given explicitly by, 
\begin{align}
	\label{eq: ref-curr}
	I^d_b=\! \! \!  \frac{V_{dc} }{V_b}\cdot \bigg(G_{dc} V_{dc}-k_P (V_{dc}-V^*_{dc})-k_I \int_{0}^t (V_{dc}(\tau)-V^*_{dc}) \dd \tau\bigg), 
\end{align}
where $k_P>0, \, k_I>0$ are proportional and integral control gains. The reference current $I^d_b$ in~\eqref{eq: ref-curr} is tracked by an inner current control loop leveraging the reference voltage $V^d_{l}:= V_b-V_c$  as follows,
\begin{align}
	\label{eq: ref-vol}
	V^d_{l} = R_b I_b -k_{BP} \big(I_b-I^d_b \big)-k_{BI} \int_{0}^t \big(I_b(\tau)-I^d_b \big) \dd \tau,
\end{align}
with $k_{BP}, \, k_{BI}>0$. The choice of the gain pairs $(k_P,k_I)$ and $(k_{BP}, k_{BI})$ ensures time-scale separation, namely that the closed-loop dynamics of the current loop is faster than that of DC voltage.
Finally, the duty cycle $d$, i.e.,  the main input to the boost converter is deduced from the reference voltage $V^d_{l}$ in~\eqref{eq: ctrl-input} and~\eqref{eq: ref-vol} via the relationship,
\begin{align}
\label{eq: DC-duty-cycle}
d= 1-\frac{V_{b} - V^d_l}{V_{dc}}.
\end{align}

\vspace{0cm}
%
\color{black}
\begin{IEEEbiography}[{\includegraphics[width=1in,height=1.25in,clip,keepaspectratio]{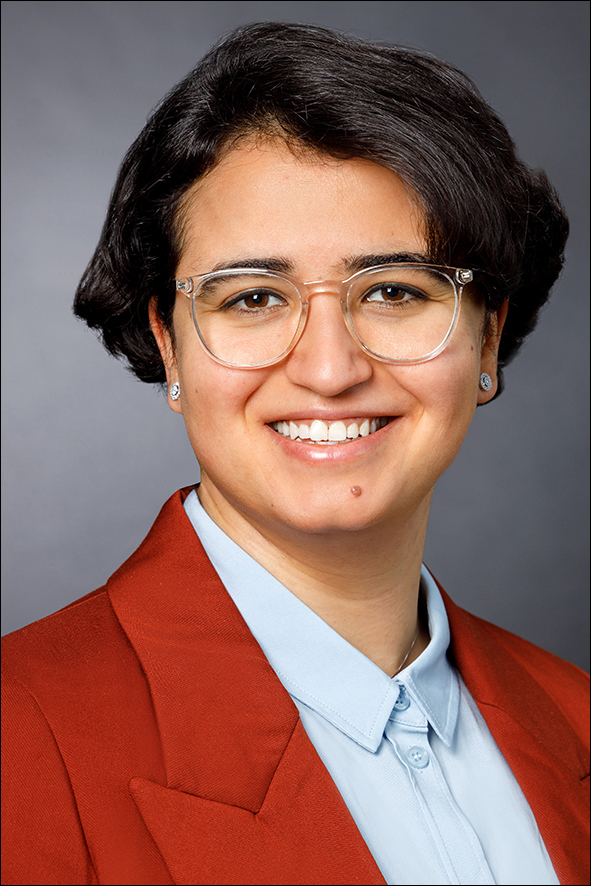}}]
{Taouba Jouini} received her PhD degree in System and Control Engineering from Lund University in 2022. She is currently a postdoctoral researcher at Leibniz University of Hannover. Her research interests are centered around the optimal management of multi-modal energy systems and the development of test environments for autonomous vehicles.
\end{IEEEbiography}
%
\begin{IEEEbiography}[{\includegraphics[width=1in,height=1.25in,clip,keepaspectratio]{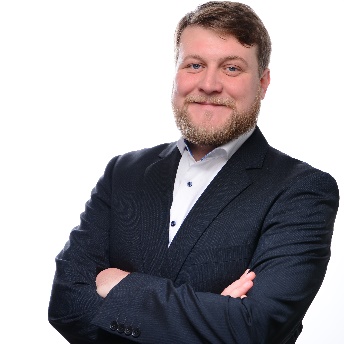}}]
{Jan Wachter} received the M.Sc. degree in mechanical engineering from the Karlsruhe Institute of Technology (KIT), Karlsruhe, Germany. He is currently working towards the Ph.d.
degree at the Institute for Automation and Applied Informatics (IAI) of the KIT. His research interest includes the control of grid connected converters systems and the challenges
arising from their interaction.
\end{IEEEbiography}
%
\begin{IEEEbiography}[{\includegraphics[width=1in,height=1.25in,clip,keepaspectratio]{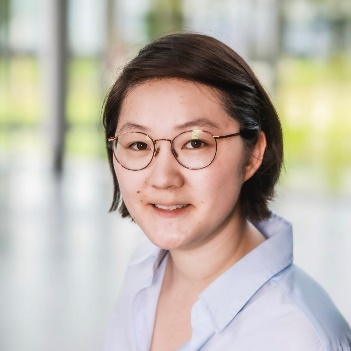}}]
{Sophie Xing An} received the M.Sc. degree in Electrical Engineering and Information Technology from the Karlsruhe Institute of Technology (KIT), Karlsruhe, Germany. She was with the Institute for Automation and Applied Informatics (IAI) of the KIT. Her research interest includes the control of the power grid in the face of the energy transition.
\end{IEEEbiography}
\begin{IEEEbiography}[{\includegraphics[width=1in,height=1.25in,clip,keepaspectratio]{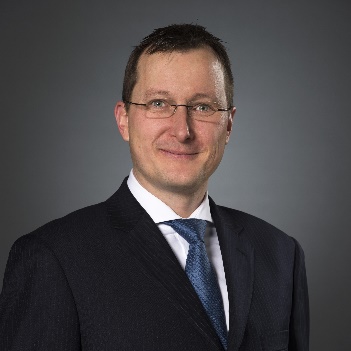}}]
{Veit Hagenmeyer} received the Ph.D. degree from Universit\'e Paris XI, Paris, France, in 2002. He is currently a Professor of energy informatics with the Faculty of Computer Science, and also the Director with the Institute for Automation and Applied Informatics (IAI), Karlsruhe Institute of Technology, Karlsruhe, Germany. His research interests include modeling, optimization and control of sector-integrated energy systems.
\end{IEEEbiography}
\vspace{15cm}
\end{document}